%% file: haw.tex
\documentstyle[12pt,epsfig,epsf]{article}
\begin{document}
\newcommand{\nn}{\noindent}
\renewcommand{\thefootnote}{\fnsymbol{footnote}}

\input fey1

\begin{flushright}
KEK/99/182\\
UFR-HEP/2000/03\\
February 2000\\
\end{flushright}

\vspace*{.4cm}
\begin{center}
{\Large\sc {\bf Radiative corrections to the decay  }}\\
\vspace*{3mm}
{\Large\sc {\bf  $H^+ \to W^+ A^0$ }} 
\vspace{.4cm}

{\large
{{A.G. Akeroyd}$^{\mbox{a,}}$\footnote{E--mail: akeroyd@post.kek.jp},
  {A. Arhrib}$^{\mbox{b,c,}}$
\footnote{E--mail: \{arhrib,naimi\}@fstt.ac.ma} and 
{E. Naimi}$^{\mbox{b,c}}$}}
\vspace{.6cm}
{\it
\\
a: KEK Theory Group, Tsukuba,\\
 Ibaraki 305-0801, Japan\\
%Fax: 0298 64 5755\\
%Tel: 0298 64 5403\\

\vspace{.6cm}
b: D\'epartement de Math\'ematiques, 
Facult\'e des Sciences et Techniques\\
B.P 416, Tanger, Morocco

\vspace{.6cm}

c: UFR--High Energy Physics, Physics Departement, Faculty of Sciences\\
PO Box 1014, Rabat--Morocco
}
\end{center}

\setcounter{footnote}{4} 
\vspace{1.2cm}

\begin{abstract}
\nn
Full one-loop electroweak corrections to the on-shell decay
$H^+ \to W^+A^0$ are computed in the framework of models with
two Higgs doublets (THDM). Such a decay may be dominant
for $H^{\pm}$ over a wide range of parameter space relevant at
present and future colliders. We show that the corrections
may approach $40\%$ and in particular are sensitive to 
$\lambda_5$, which parametrizes the discrete symmetry 
breaking term. We suggest that a measurement of the branching ratio of 
$H^+ \to W^+A^0$ may offer a possibility of measuring the
magnitude of $\lambda_5$.

\end{abstract}
\vspace{1.2cm}


%hep-ph/0002288
\newpage
\pagestyle{plain}
\renewcommand{\thefootnote}{\arabic{footnote} }
\setcounter{footnote}{0}

\section*{1.~Introduction}
The phenomenology of charged Higgs bosons ($H^{\pm}$)
has received much attention
in recent years \cite{Gun} since their discovery would 
provide conclusive evidence of physics beyond the Standard Model (SM) 
\cite{Wein}. Charged Higgs bosons are predicted in many theoretically
well-motivated extensions of the SM. The simplest model which
contains a $H^{\pm}$ is the Two Higgs 
Doublet Model (THDM), which is formed by adding an extra complex
$SU(2)_L\otimes U(1)_Y$ scalar doublet to the SM lagrangian. 
Motivations for such a structure include CP--violation in the Higgs 
sector and a possible solution to the cosmological 
domain wall problem \cite{preskill}. In particular, the Higgs
sector of the Minimal 
Supersymmetric Standard Model (MSSM) \cite{Gun} takes the form
of a constrained THDM.

The phenomenology of $H^{\pm}$ has received substantial attention at 
$e^+e^-$ colliders
\cite{e+e-},\cite{ArhribMoult}, hadron colliders \cite{Hadron}, 
\cite{recent}, \cite{progress},
$\mu^+\mu^-$ colliders \cite{muon} and $\gamma\gamma$ colliders \cite{yy}. 
Most phenomenological studies have been carried 
out in the  context of the MSSM.
The combined null--searches from all four CERN LEP collaborations derive the 
lower limit $m_{H^{\pm}}\ge 77.4$ GeV $(95\%\, c.l)$ \cite{LEP}, a limit
which applies to all models in which BR($H^{\pm}\to \tau\nu_{\tau}$)+
BR($H^{\pm}\to cs$)=1, where BR signifies branching ratio.
Current mass bounds from LEP--II for the neutral pseudoscalar $A^0$ 
of the MSSM 
($m_A\ge 90.5$ GeV) force $m_{H^{\pm}}\ge 120$ GeV 
in this model \cite{Roy}, which is stronger than the direct search limit 
above. Limits on $m_{H^{\pm}}$ from the Fermilab Tevatron searches \cite{tev}
are $\tan\beta$ dependent since 
a significant BR$(t\to H^{+}b$) is required in order to obtain a 
visible signal.
The limits are competitive with those from LEP--II for the regions 
$\tan\beta\le 1$ or $\ge 40$.

In the MSSM, $m_{H^{\pm}}$ and the mass of the pseudoscalar 
$m_A$ are 
approximately degenerate for values greater than 200 GeV, and so the 
two body decay 
$H^{\pm}\to A^0W$ is never allowed for masses of interest at the 
CERN Large Hadron Collider (LHC).
The three--body decay $H^{\pm}\to A^0W^*\to A^0f\overline f$ is open
for smaller $m_{H^{\pm}}$ although it possesses a small branching ratio
(BR$\le 5\%$ for $m_{H^\pm}\ge 110$ GeV) \cite{djou3body}.
Therefore for masses of interest at the LHC the principal decay channel is 
$H^{\pm}\to tb$, with decays to SUSY particles possibly open as $m_{H^{\pm}}$
increases \cite{djou}. Recently there has been a surge of interest
in  studies of the MSSM with unconstrained CP--violating phases. In such
scenarios $H^0-A^0$ mixing may be induced radiatively \cite{CP} although this
only leads to maximum mass splittings of the order 20 GeV for small values of
$\tan\beta$; thus the two body decay would not be open even in this
case.

Non--supersymmetric THDMs (hereafter to be 
called simply 'THDM') 
have also received considerable attention in the literature.
In such models all the Higgs masses may be taken as free parameters 
(in contrast to the MSSM), 
thus allowing the possibility of the two body decay $H^{\pm}\to A^0W$ 
for certain choices of $m_A$ and $m_{H^{\pm}}$. 
This decay mode posesses no mixing angle suppression, in contrast to
$H^{\pm}\to h^0W$, and may compete with conventional decays
\cite{3body},\cite{Bor}. In fact, 
in a sizeable region of parameter space we show that it may be 
the dominant channel. Motivated by the results
in Ref.~\cite{3body} the authors calculated in Ref. \cite{HAW} the
Yukawa corrections to the decay $H^{\pm}\to A^0W^{(*)}$ 
and found that in the on-shell case the corrections may 
approach 50\%  for small values of $\tan\beta$. 
In this paper we complete that analysis and
include the full bosonic corrections for the case of 
the $W$ being  on-shell.

Conventional Higgs searches at LEP--II assume the decays
$H^{\pm}\to \tau\nu_{\tau}$ and $cs$ \cite{LEP}, 
although the OPAL collaboration has recently carried out
the first search for $H^{\pm}\to A^0W^{*}$ topologies 
\cite{Hoffman} in the context of the THDM (Model~I).
A recent study \cite{hawsig} showed that the decay 
$H^{\pm}\to A^0W$ offers good 
chances of detection for $H^{\pm}$ at the LHC, where an analysis in the 
context of the MSSM with an extra singlet superfield was carried out (NMSSM).
Much of this work would be relevant for the THDM that we consider, 
although our bosonic corrections would not be directly applicable to the 
NMSSM, since the latter possesses two more neutral Higgs bosons 
in addition to different Higgs self-couplings. 

The paper is organized 
as follows. In section 2 we introduce our notation and outline the
form of the 1--loop corrections. In section 3 we explain the 
importance of including diagrams with the emission of a 
soft photon, $H^+\to W^+A^0\gamma$, in order to keep the radiative
corrections infra-red finite. Section 4 covers the various experimental
and theoretical constraints that we impose. Section 5 presents our
numerical results for the full corrections (bosonic and Yukawa), while
section 6 contains our conclusions. The explicit form of the corrections
is contained in the appendix.

\renewcommand{\theequation}{2.\arabic{equation}}
\setcounter{equation}{0}
\section*{2. Lowest order result and structure of one-loop radiative corrections}
\subsection*{2.1 Lowest order result}
We will be using the notation and conventions of
our previous work \cite{HAW}, which we briefly review here. The 
momentum of the charged
Higgs boson $H^+$ is denoted by $p_H$ ($p_H$ is incoming),
${p_W}$ is the momentum of the $W^+$ gauge boson and $p_A$ the 
momentum of the CP-odd $A^0$ ($p_W$ and $p_A$ are outgoing). 
\par
The relevant part of the lagrangian describing the interaction 
of the $W^\pm$ with $H^\pm$ and $A^0$ comes from the covariant 
derivative which is given by:
\begin{eqnarray}
{\cal L} = \frac{e}{2 s_W} W_{\mu}^+ 
( H^- \stackrel{\leftrightarrow}{\partial}^{\mu} A^0) +
\mbox{h.c.} \label{lag}
\end{eqnarray}
This interaction is model independent (SUSY or non--SUSY) and depends 
only on standard parameters: electric charge ($e$) and Weinberg angle  
($s_W=\sin\theta_W$).

The lowest--order Feynman diagram for the two--body decay 
$H^+\to A^0 W^+$ is depicted in the following figure:\
\begin{center}
\begin{picture}(100,100)(0,0)
\put(60,50) {\unbs{$W^+$}}
\put(60,50) {\unfsp{$A^0$}}
\multiput(9,50)(8,0){7}{\line(1,0){6}}
\put(5,54){\mbox{$H^+$}}
\put(64,50){\circle*{5}}
\end{picture}
\end{center}
\vspace{.1cm}
{\centerline{\bf Figure. 1}}
\vspace{.4cm}
In the Born approximation, the decay amplitude of the charged Higgs into 
an \underline{on-shell} CP--odd Higgs
boson $A^0$ and the gauge boson $W^+$ (Fig.1) 
can be written as:
\begin{eqnarray}
& & {\cal M}^0 (H^+ \to W^+ A^0 ) = 
\epsilon^*_{\mu} \Gamma_0^{\mu} \label{amp0}
\qquad \mbox{where} \qquad
\Gamma_0^{\mu} = i\frac{e}{2 s_W} (p_H + p_A)_{\mu}
\end{eqnarray}
Here $\epsilon_{\mu}$ is the $W^{\pm}$ polarization vector.
We then have the following decay width:
\begin{eqnarray}
& & \Gamma_{on}^0 =
\frac{\alpha}{16 s_W^2 m_W^2 m_{H^{\pm}}^3 } 
\lambda^{\frac{3}{2}}(m_{H^{\pm}}^2, m_{A}^2,m_W^2 ) \label{width0}
\end{eqnarray}
where $\lambda=\lambda(x,y,z)=x^2+y^2+z^2 -2(xy+ xz+ yz)$ 
is the familiar two--body phase space function.
Note that in the MSSM the two--body decay of the charged
Higgs boson into $W^+A^0$ is kinematically not allowed.
In this paper we will not present results for the case of $W^{\pm}$ 
being off--shell. Ref.~\cite{HAW} evaluated the Yukawa corrections 
in the off--shell case for $m_{H^{\pm}}$ in the range of LEP--II, 
finding maximum values of a few percent. 

\subsection*{2.2 One--Loop radiative corrections}
We shall evaluate the bosonic one-loop radiative corrections 
to the decay $H^+ \to W^+ A^0$, and add them to the Yukawa corrections 
previously evaluated in Ref.~\cite{HAW}.
This set of  corrections is ultra--violet (UV) and infra--red (IR) 
divergent. The UV singularities are treated by dimensional 
regularization \cite{thooft} in the on--mass--shell renormalization scheme.
The IR divergences are treated by the introduction of a small 
fictitious mass $\delta$ for the photon, which we shall 
explain in the next section.

The typical Feynman diagrams for the virtual corrections of order $\alpha$ 
are listed in figure 2.1 $\to$2.16. These contributions have to be 
supplemented by 
the counterterm renormalizing the vertex $H^+A^0W^- $  (eq \ref{dlag}).
Note that in the THDM, the vertices $W^+A^0 G^-$, 
$W^+G^0 H^-$, $W^+W^- A^0$  and $A^0 H^+ H^-$ are not present, and so the 
mixing $G^+$--$H^+$, $G^0$--$A^0$ and $W^+$--$H^+$ does not give any 
contribution to our process. In our case,
the gauge boson $W$ is on-shell and so the mixing $W^\pm$-$G^\mp$ is absent.
The full set of Feynman diagrams are generated and computed
using the FeynArts and FeynCalc \cite{seep,rolf} packages. 
The amplitudes
of the typical vertices are given in terms of the one-loop 
scalar functions \cite{passarino} and are written explicitly in 
appendix B. We also use the fortran FF--package \cite{ff} 
in the numerical analysis.

In what follows we will use the on-shell
renormalization scheme developed in Ref.~\cite{HAW} 
(and refs therein). The vertex counterterm is given by:
\begin{eqnarray}
\delta {\cal L} = 
\frac{e}{2 s_W} W_{\mu}^+ (H^- \stackrel{\leftrightarrow}{\partial}^{\mu} 
A^0 )( \frac{1}{2} \delta Z_{WW} +
\frac{1}{2} \delta Z_{A^0A^0}+
\frac{1}{2} \delta Z_{H^{\pm}H^{\pm}} +
\delta Z_e - \frac{\delta s_W}{s_W} )  \label{dlag}
\end{eqnarray}
where
$\delta Z_{WW}$, $\delta Z_{A^0A^0}$ and $\delta Z_{H^{\pm}H^{\pm}}$ are
the wave function renormalization constants for the $W^{\pm}$ gauge boson,
$A^0$ and $H^\pm$ Higgs boson defined as follows:
\begin{eqnarray}
& & \delta Z_{ii} = -\frac{\partial\Sigma_{ii}
(k^2)}{\partial k^2} |_{k^2=m_i^2}\label{dmz}  \ \ \ i=W, \ A^0 \ H^\pm 
\label{l1}\\ 
& & \delta m_i^2 = Re \, {\Sigma_{ii} } (m_i^2) \ \ \ \ \ \ i=W, \ Z
\label{l2}
\end{eqnarray}
where $\Sigma_{ii} (k^2)$ is the  bare self-energy of the $H^{\pm}$, 
$A^0$ or  $W$. The electric charge counterterm and 
$\frac{\delta s_W}{s_W}$ 
are defined as:
\begin{eqnarray}
& & \frac{\delta s_W}{s_W}  =  -\frac{1}{2} \frac{c_W^2}{s_W^2} 
( \frac{\delta m_W^2}{m_W^2} - \frac{\delta m_Z^2}{m_Z^2} )\label{l3}\\  
& & \delta Z_e = -\frac{1}{2} \delta Z_{\gamma\gamma} +
\frac{1}{2} \frac{s_W}{c_W} \delta Z_{Z\gamma} = 
\frac{1}{2} \frac{\partial\Sigma_T^{\gamma\gamma}(k^2)}{\partial k^2}
\vert_{k^2=0} + \frac{s_W}{c_W}  
\frac{\Sigma_T^{\gamma Z}(0)}{m_Z^2} \label{dmv2}
\end{eqnarray}
The index $T$ in $\Sigma_T^{\gamma\gamma}$ and 
$\Sigma_T^{\gamma Z}$
denotes that we take the transverse part. The scalar one-loop self-energies
entering in the above equations (\ref{l1}$\to$ \ref{dmv2}) are given in 
appendix $C_1$ and $C_2$, while gauge boson self energies can be 
found in \cite{gauge}.

The one-loop amplitude ${\cal M}^1$ (vertex plus counterterms)  
can be written as:
\begin{eqnarray}
& & {\cal M}^1 (H^+ \to W^+ A^0 ) = 
\frac{e}{2 s_W}(\Gamma_{H} p_H^{\mu} +\Gamma_{W} p_W^{\mu}) 
 \epsilon^*_{\mu} \label{amp1}
\end{eqnarray}
where $\Gamma_{H}$ and $\Gamma_{W}$ can be cast as follow:
\begin{eqnarray}
 & & \Gamma_{W}= \Gamma_{W}^{vertex} +
 \delta\Gamma_{W}^{vertex}  \\
 & & \Gamma_{H}= \Gamma_{H}^{vertex} +  
\delta\Gamma_{H}^{vertex} \label{forms}
\end{eqnarray}
Here $\Gamma_{W,H}^{vertex}$ 
represents the vertex corrections and $\delta\Gamma_{W,H}^{vertex}$
is the counterterm contribution needed to remove the UV divergences
contained in $\Gamma_{W,H}^{vertex}$. 

The expressions of the counterterms are: 
\begin{eqnarray}
& & \delta\Gamma_{W}^{vertex} = -(\delta Z_e -\frac{\delta s_W}{s_W} +
\frac{1}{2} (\delta Z_{H^+H^+} + \delta Z_{A^0} + \delta Z_{W}) )\nonumber\\
& &   \delta\Gamma_{H}^{vertex} =
-2 \delta\Gamma_{W}^{vertex}
\end{eqnarray}
In the on-shell case the interference term 
$2Re{\cal M}^{0*}{\cal M}^1$, found from squaring 
the one-loop corrected amplitude 
$|{\cal M}^0+{\cal M}^1|^2$, is equal to 
$\Gamma_H|{\cal M}^0|^2$ \cite{HAW}.
Hence the one-loop corrected width $\Gamma^1_{on}$ 
can be written as
\begin{equation}
\Gamma^1_{on}=(1+\Gamma_H)\Gamma^0_{on} 
\end{equation}
with $\Gamma_H$ (defined by eq. \ref{forms}) being interpreted as 
the fractional contribution to the tree-level width, $\Gamma^0_{on}$.
Note that $\Gamma_W$ (eq. 2.10) does not 
contribute to $\Gamma^1_{on}$.
\renewcommand{\theequation}{3.\arabic{equation}}
\setcounter{equation}{0}

\section*{3. Real photon emission: $H^+\to W^+ A^0\gamma$}
The vertex correction supplemented by the counterterms is UV finite
but there still remains infra-red divergences. These arise from the
diagrams 2.10 and 2.11 with $V=\gamma $ and also from 
the wave function renormalisation constant $\delta Z_{H^\pm H^\pm }$
(Diagram 2.33) and 
$\delta Z_{WW}$. 
In order to obtain a finite result, one has to add the correction from 
the emission of a real photon in the final state as drawn in figures
$2.17\to 2.19$. 

In terms of the momenta of the particles in the final state, 
the square amplitude of the process $H^+ \to A^0 W^+ \gamma$ is given by:
\begin{eqnarray}
|M(H^+\to A^0 W^+ \gamma)|^2 & = & -\frac{e^4}{s_W^2}
\lambda(m_{H{\pm}}^2,m_A^2,m_W^2)
[\frac{m_{H{\pm}}^2}{m_W^2}\frac{1}{(2 p_Hk_\gamma)^2}\nonumber\\ &&
+\frac{1}{(2 p_W k_\gamma)^2} + (\frac{m_A^2-m_{H^{\pm}}^2}{m_W^2}-1)
\frac{1}{(2 p_W k_\gamma)(2 p_H k_\gamma)}]
\end{eqnarray}
where $k_\gamma$ denotes the momentum of the photon. Note that as a 
consequence of gauge invariance, 
the amplitude of the sum of the three diagrams (Figs. $2.17\to 2.19$), 
should vanish when multiplied by
the four-momentum of the photon, which provides a good check of the 
calculation. The integrals over three body phase space can be found in 
\cite{denner}, and one obtains the following expression for the
width:
\begin{eqnarray}
\Gamma_{Br} & = & -\Gamma_{on}^0 \frac{e^2 M^2_{H^\pm}}
{ \pi^2\lambda^{\frac{1}{2}}}
[\frac{m_{H{^\pm}}^2}{m_W^2}I_{HH}
+I_{WW} +(1+\frac{m_{H{^\pm}}^2-m_A^2}{m_W^2})I_{HW}]
\end{eqnarray}
Where $I_{HH}$ , $I_{WW}$ and $I_{HW}$ 
are given as follows:
\begin{eqnarray}
&& I_{HH}=\frac{1}{4 m_{H^\pm}^4}\{ 
\lambda^\frac{1}{2}
\mbox{log}(\frac{\lambda }
{\delta m_{H^\pm}m_Am_W}) - 
\lambda^\frac{1}{2}
-(m_W^2 - m_A^2) \mbox{log}(\frac{\beta_1}{\beta_2})
-m_{H^\pm}^2\mbox{log}(\beta_0)\}\nonumber \\
& & I_{WW}=\frac{1}{4 m_{H^\pm}^2m_W^2}\{ 
\lambda^\frac{1}{2}
\mbox{log}(\frac{\lambda }
{\delta m_{H^\pm}m_Am_W}) - 
\lambda^\frac{1}{2}
-(m_{H^\pm}^2 - m_A^2) \mbox{log}(\frac{\beta_0}{\beta_2})
-m_{W}^2\mbox{log}(\beta_1)\}\nonumber \\
& & I_{HW}=\frac{1}{4 m_{H^\pm}^4}\{ 
2 \mbox{log}(\frac{\lambda }
{\delta m_{H^\pm}m_Am_W})\mbox{log}(\beta_2) + 2 \mbox{log}^2(\beta_2)-
\mbox{log}^2(\beta_0)
-\mbox{log}^2(\beta_1) \nonumber \\ & & \qquad \qquad + 2 \mbox{Sp}(1-\beta_2^2)
- \mbox{Sp}(1-\beta_0^2) - \mbox{Sp}(1-\beta_1^2)\}
\end{eqnarray}
where $\lambda=\lambda (m_{H^\pm}^2, m_A^2,m_W^2)$ is the two body phase
space, $\delta$ is a small fictitious photon mass, 
$Sp$ is the dilogarithm function  and $\beta_i$ are
defined as:
\begin{eqnarray}
& & \beta_0=\frac{m_{H^\pm}^2-m_W^2-m_A^2 +\lambda^{\frac{1}{2}} }{2 m_W m_A} 
\qquad , \qquad
\beta_1=\frac{m_{H^\pm}^2-m_W^2+m_A^2 -\lambda^{\frac{1}{2}} }{2 m_{H^\pm} m_A}
\qquad , \qquad \nonumber \\ & &
\beta_2=\frac{m_{H^\pm}^2+m_W^2-m_A^2 -\lambda^{\frac{1}{2}} }{2 m_{H^\pm} m_W}
\end{eqnarray}
We stress here that the IR divergence contained in $I_{HH}$ ($I_{WW}$) 
is cancelled by the wave function renormalisation constant of the charged Higgs 
$H^\pm$ ($W^\pm$), while the IR 
divergence contained in $I_{HW}$ is cancelled by the vertex 
diagrams 2.10 and 2.11 (with $V=\gamma$). One can confirm easily that 
adding the virtual corrections with the Bremsstrahlung diagrams yields 
an IR finite result. This feature 
has been checked both algebraically and numerically.

\renewcommand{\theequation}{4.\arabic{equation}}
\setcounter{equation}{0}
\section*{4. THDM scalar potential: Theoretical and 
Experimental constraints}
In this section we define the THDM scalar potential that we will be using. 
In appendix A we list the trilinear and quartic scalar 
self--couplings which are relevant to our study. 
Other relevant couplings involving Higgs boson interactions with 
gauge bosons and fermions can be found 
in Ref.~\cite{Gun}. For a full list of scalar trilinear 
and quartic couplings see Ref.~\cite{Thesis}. It has been shown 
\cite{Gun} that the most general THDM scalar potential which is both 
$SU(2)_L\otimes U(1)_Y$ and CP invariant is given by:
\begin{eqnarray}
 V(\Phi_{1}, \Phi_{2})& & =  \lambda_{1} ( |\Phi_{1}|^2-v_{1}^2)^2
+\lambda_{2} (|\Phi_{2}|^2-v_{2}^2)^2+
\lambda_{3}((|\Phi_{1}|^2-v_{1}^2)+(|\Phi_{2}|^2-v_{2}^2))^2 
+\nonumber\\ [0.2cm]
&  & \lambda_{4}(|\Phi_{1}|^2 |\Phi_{2}|^2 - |\Phi_{1}^+\Phi_{2}|^2  )+
\lambda_{5} (Re(\Phi^+_{1}\Phi_{2})
-v_{1}v_{2})^2+ \lambda_{6} [Im(\Phi^+_{1}\Phi_{2})]^2 
\label{higgspot}
\end{eqnarray}
where $\Phi_1$ and $\Phi_2$ have weak hypercharge Y=1, $v_1$ and
$v_2$ are respectively the vacuum
expectation values of $\Phi_1$ and $\Phi_2$ and the $\lambda_i$
are real--valued parameters. 
Note that this potential violates the discrete symmetry
$\Phi_i\to -\Phi_i$ softly by the dimension two term
$\lambda_5 Re(\Phi^+_{1}\Phi_{2})$ and has the same 
general structure as the scalar potential of the MSSM.
One can prove easily that for $\lambda_5=0$  
the exact symmetry $\Phi_i \to -\Phi_i$ is recovered.
We note that Ref.~\cite{santos} 
lists the complete Higgs trilinear and quartic interactions 
for two 6 parameter potentials,
referred to as 'Potential A' and 'Potential B'. Potential A 
is equivalent to our potential if $\lambda_5\to 0$, and in 
this limit the Feynman rules in the appendix A are in agreement 
with those in Ref.~\cite{santos}.

After electroweak symmetry breaking, the W and Z gauge 
bosons acquire masses  
given by  $m_W^2=\frac{1}{2}g^2 v^2$ and 
$m_Z^2= \frac{1}{2}(g^2 +g'^2) v^2$,
where $g$ and $g'$ are the $SU(2)_{L}$ and 
$U(1)_Y$ gauge couplings and
$ v^2= v_1^2 + v_2^2$. The combination $v_1^2 + v_2^2$ 
is thus fixed by the electroweak 
scale through $v_1^2 + v_2^2=(2\sqrt{2} G_F)^{-1}$, 
and we are left with 7 free parameters in eq.(\ref{higgspot}), 
namely the $(\lambda_i)_{i=1,\ldots,6}$ and 
$\tan\beta=v_2/v_1$. Meanwhile,  three of the eight degrees of freedom 
of the two Higgs doublets correspond to 
the 3 Goldstone bosons ($G^\pm$, $G^0$) and  
the remaining five become physical Higgs bosons: 
$H^0$, $h^0$ (CP--even), $A^0$ (CP--odd)
and $H^\pm$. Their masses are obtained as usual 
by the shift $\Phi_i\to \Phi_i + v_i$ and read \cite{Gun}: 
\begin{eqnarray}
&& m_A^2=\lambda_6 v^2\quad , \quad m_{H^{\pm}}^2=
\lambda_4 v^2\nonumber \quad , \quad 
m_{H,h}^2=\frac{1}{2} [ A+C \pm \sqrt{(A-C)^2+4B^2} ]
\label{higgsmass}
\end{eqnarray}
where
\begin{eqnarray}
& & A=4 v_1^2 (\lambda_1+\lambda_3)+v_2^2\lambda_5\ , \ 
 B= v_1 v_2 (4\lambda_3+\lambda_5)\ \mbox{and} \
C=4 v_2^2 (\lambda_2+\lambda_3)+v_1^2\lambda_5
\label{ABC}
\end{eqnarray}
The angle $\beta$ diagonalizes both the CP--odd and charged scalar
mass matrices, leading to the physical states $H^\pm$ and $A^0$.
The CP--even mass matrix is diagonalized by the angle $\alpha$, leading
to the physical states $H^0$, $h^0$, with  $\alpha$ given by:
\begin{eqnarray}
&& \sin 2 \alpha=\frac{2 B}{\sqrt{(A-C)^2+4 B^2} }\ , \ \ \
\cos 2 \alpha=\frac{A-C}{\sqrt{(A-C)^2+4 B^2} }
\label{alph}
\end{eqnarray}
It is then straightforward algebra to invert the previous equations
to obtain the $\lambda_i$ in terms of physical scalar masses, 
$\tan\beta$, $\alpha$ and $\lambda_5$:
\begin{eqnarray}
& & \lambda_4=\frac{g^2}{2 m^2_W} m_{H^\pm}^2 \ \ ,  \ \
\lambda_6=\frac{g^2}{2 m^2_W} m_{A}^2  \label{lambda46} 
\ \ , \ \  \lambda_3=\frac{g^2}{8 m^2_W}
\frac{\mbox{s}_\alpha \mbox{c}_\alpha}{ \mbox{s}_\beta \mbox{c}_\beta } 
(m_H^2-m_h^2)\ -
\  \frac{\lambda_5}{4} \label{lambda5}  \\
& & \lambda_1 = \frac{g^2}{8 \mbox{c}_\beta^2 m^2_W} 
[ \mbox{c}_\alpha^2 m^2_H+
\mbox{s}_\alpha^2 m^2_h -
\frac{\mbox{s}_\alpha \mbox{c}_\alpha}{\tan\beta}(m^2_H - m^2_h)] 
 -\frac{\lambda_5}{4}(-1 + \tan^2\beta) \label{lambda1} \\ & &
\lambda_2 = \frac{g^2}{8 \mbox{s}_\beta^2 m^2_W} [ \mbox{s}_\alpha^2 m^2_H+
\mbox{c}_\alpha^2 m^2_h -
\mbox{s}_\alpha \mbox{c}_\alpha \tan\beta(m^2_H - m^2_h)] 
 -\frac{\lambda_5}{4}(-1 + \frac{1}{\tan^2\beta} ) \label{lambda1} 
\label{lambda2}
\end{eqnarray}
We are free to take as 7 independent parameters 
$(\lambda_i)_{i=1,\ldots , 6}$ and $\tan\beta$
or equivalently the four scalar masses, $\tan\beta$, $\alpha$
and one of the $\lambda_i$. In what 
follows we will take $\lambda_5$ as a free parameter.
In our analysis we also take into account the following 
constraints when the independent parameters are varied.
\\
$\bullet$ The contributions to the $\delta\rho$ parameter from the Higgs
scalars \cite{Rhoparam} should not exceed the current limits from precision 
measurements \cite{Langacker}: $ -0.0017 \leq \delta\rho \leq 0.0027$.\\
$\bullet$ From the requirement of perturbativity for the
top and bottom Yukawa couplings \cite{berger}, $\tan\beta$ is 
constrained 
to lie in the range $0.3\leq \tan\beta \leq 130$. Upper and lower 
bounds have also been obtained from the experimental limits on the processes 
$e^+e^-\to Z^*\to h^0\gamma$ and/or $e^+e^-\to A^0\gamma$. 
For very light $h$ or $A^0$ ($\approx 10$ GeV)
Ref.~\cite{kraw} derived $0.15\leq \tan\beta \leq 75$, with the 
limits weakening for heavier $m_h(m_A$).
For our study we will restrict the discussion to values 
$\tan\beta\geq 0.5$.\\
$\bullet$ We require that tree-level unitarity is not violated in a variety
of Higgs scattering processes \cite{prepar}.

\renewcommand{\theequation}{5.\arabic{equation}}
\setcounter{equation}{0}
\section*{5. Numerical results and discussion}
In this section we present our numerical results for $\Gamma_H$, 
which is the fractional correction to the tree-level width (eq.2.13). 
We take the following experimental input for the physical parameters 
\cite{databooklet}. The fine structure constant: 
$\alpha=\frac{e^2}{4\pi}=1/137.03598$, 
the gauge boson masses: $m_Z=91.187\ GeV$, $m_W=80.41\ GeV$, the 
lepton masses: $m_e=0.511 \mbox{MeV}$, $ m_{\mu}=0.1057 \mbox{GeV}$, 
$m_{\tau}=1.784  \mbox{GeV} $. For the light quark masses 
we use the effective values which are chosen in 
such a way that the experimentally extracted hadronic part of the 
vacuum polarizations is reproduced 
\cite{martinzepenfieldVerzeganssijegerlener}:
$m_d=47\ MeV \ ,\ m_u=47 \ MeV \ , \ m_s=150\ MeV \ , \ m_c=1.55\ GeV \ , \ 
m_b=4.5 \ GeV $.
For the top quark mass we take $m_t=175$ GeV. 
In the on-shell scheme we consider, $\sin^2 \theta_W$ is given by
$\sin^2 \theta_W\equiv 1- \frac{m_W^2}{m_Z^2}$, and this
expression is valid beyond tree-level.\\

Let us make some comment about the Yukawa corrections discussed 
in Ref.~\cite{HAW}. In the case of an on-shell $W$, we showed 
that for small $\tan\beta$  in both Model~I and II
one can find large corrections of up to around $\pm 10\%$
(away from threshold effects at $m_A\approx 2m_t$). In Model I for large 
$\tan\beta$ all fermion corrections decouple and reach a constant value 
of 3.3\% for $\tan\beta>4$. In Model II 
$\Gamma_H$ is enhanced for large $\tan\beta$
($> 20$) since in this scenario the internal $b$ quarks in the 
loop couple more strongly to $A^0$.
Typically for $m_{H^\pm}=440$ one can reach a correction 
of about $-7\% \to -20\%$ for large $\tan\beta >90$ and
light $m_A$ ($60 \leq m_A\leq 100$ GeV).  For $m_A>100$ GeV 
the $\tan\beta$ dependence of the 
Yukawa correction is rather weak and lies in the range of 
$-5\% \to 5\% $ for $100 \leq m_A\leq 330$ GeV.

We stress at this stage that in the low $\tan\beta$ ($\tan\beta<2$) regime 
the corrections in Model I and Model II are practically identical. 
Perturbative constraints on the $\lambda_i$ and 
unitarity constraints on the quartic scalar couplings constrain 
the magnitude of $\tan\beta$. As shown in \cite{prepar},
$\tan\beta\ge 20$ violates the unitarity bounds 
if $\lambda_5=0$, although for $\lambda_5\ne 0$ values of
$\tan\beta\ge 40$ are comfortably allowed. 
However, perturbative constraints on the $\lambda_i$ (in particular 
$\lambda_1$) disfavour $\tan\beta\ge 30$ \cite{prepar}, and so in our 
analysis we will only consider small to moderate values of $\tan\beta$. 
Therefore our results are applicable to both Model I and II.
Note that in order to satisfy the experimental constraint on 
$\delta\rho$ we have assumed (for the graphs we have plotted)
that $\alpha=\beta-\frac{\pi}{2}$ and the charged Higgs boson mass
$m_{H^\pm}$ is quasi-degenerate with $m_H$.

In Fig.3 we plot $\Gamma_H$ as a function of 
$m_{H^\pm}$ for $m_H=m_{H^\pm}-10$ GeV, $m_h=120$ GeV, 
$m_A=150$ GeV and $\alpha=\beta-\frac{\pi}{2}$ for 
several values of $\lambda_5$. 
With the above set of parameters and for
$\tan\beta=0.5$(1.5) and $\lambda_5=0$ in Fig.3.a (Fig.3.b),   
the unitarity constraints in the spirit of Ref. \cite{prepar}
require $m_{H^\pm}\le 370$
(480) GeV, with this bound weakening for increasing $\lambda_5$.
This can be seen both in fig.3.a and fig.3.b, 
where we cut the curves at the value of the charged Higgs mass
which violates the unitarity constraint.
Both for  Fig.3.a and Fig.3.b the Yukawa correction is positive and 
lies in the range $3.6\% \to 11.9\% $ and $3.3\% \to 4.3\% $ for 
Fig.3.a and Fig.3.b respectively 
while the bosonic correction is negative.  
In Fig.3.a ($\tan\beta=0.5$), the bosonic correction is in 
the range $-0.6\%\to -0.9\%$ for 
$\lambda_5=0$ and $m_{H^\pm}\in [231,370]$, and so in 
this case the Yukawa correction is dominant;
for larger $\lambda_5=8.5$ the bosonic correction becomes strongly 
negative ( $-11\%\to -50\%$ for  $m_{H^\pm}\in [230,620]$)
and dominates the Yukawa corrections.

In Fig.3.b we take $\tan\beta=1.5$, 
and the bosonic correction is in the range $ -0.6\% \to -1.6\% $ 
for $\lambda_5 =0$ and $m_{H^\pm}\in [230, 475] $; for
$\lambda_5=1$ and $ m_{H^\pm}\in [230,520]$ it is in the range
$ -1.7\% \to - 4.7\% $ . In the above two cases
the bosonic and Yukawa corrections interfere destructively
leading to a small total correction of about $\approx 3\%$.
In the case where $\lambda_5\geq 3$ the bosonic correction 
becomes strongly negative and dominates the Yukawa correction.

Fig.4.a and Fig.4.b show the total contribution to $\Gamma_H$ 
as function of $m_A$ for $m_H=500$, $m_h=360$ and $m_{H^\pm}=530$ GeV. 
Fig.4.a corresponds to $\tan\beta=0.8$
and Fig.4.b corresponds to $\tan\beta=1.6$
(with $\alpha=\beta -\frac{\pi}{2}$).
In both figures the Yukawa correction is positive in the 
region $m_A\le 277$ GeV and $m_A\ge 352$ GeV,
while for $m_A \approx 350$ GeV the channel $A\to tt$ opens, leading to 
a very large negative correction. Note that the bosonic correction is 
negative for every value of $m_A$.

We can conclude that for the intermediate mass range of $m_A$ (away from 
threshold effects $m_A\approx 2 m_t$)
and for $ \lambda_5 \leq 4 $  
there is a cancellation between the Yukawa correction and the 
bosonic correction. For large $\lambda_5$ the contribution to 
$\Gamma_H$ is dominated by 
the bosonic correction and is consequently negative. For 
$m_A \approx 350$ GeV ($\approx 2m_t$) there is constructive interference 
between the Yukawa and bosonic corrections.

In Fig.5.a we plot the bosonic contribution to $\Gamma_H$ 
(denoted $\Gamma_H^{bos}$) as a function 
of $\lambda_5$ for $\tan\beta=0.5,5,8$. 
The fermionic correction (independent of $\lambda_5$) takes the
values -3.21\%, 3.22\%, 3.24\%. 
One can see from the curves that $\Gamma_H^{bos}$ increases
with $\lambda_5$ and may approach $-40\%$.
This is expected from the form of the trilinear scalar couplings which
increase linearly with $\lambda_5$.

Fig.5.b shows the dependence of $\Gamma_H^{bos}$ on $\sin\alpha$
for $-\pi/2<\alpha<\pi/2$. Here $C_1$, $C_2$ and $C_3$ correspond to 
three distinct parameter configurations which satisfy both
the unitarity and $\rho$ parameter constraints (see figure caption).
The fermionic corrections (independent of $\alpha$) for 
$C_1$, $C_2$ and $C_3$ take the values 3.2\% , 3.5\% and 3.4\% respectively.
The bosonic corrections become important for $\sin\alpha\approx \pm 1$.

It is apparent from the figures that $\Gamma^{bos}_H$ has a complicated 
dependence on $\alpha$, $\beta$, $\lambda_5$ and the physical Higgs masses.
This is clear from the explicit form of the trilinear couplings
in appendix A.1, which mediate the numerous triangular loop corrections.
Therefore enhancement in the bosonic sector may occur in a variety
of scenarios. Of particular interest is the sensitivity to $\lambda_5$,
a measurement of which (along with the Higgs masses and mixing angles) would
allow reconstruction of the Higgs potential. We suggest the measurement
of BR$(H^{\pm}\to A^0W)$ as a way of obtaining information on $\lambda_5$.
The decay mode $(H^{\pm}\to A^0W$) may in fact be dominant and in Fig.6 
we show the ratio 
\begin{eqnarray}
R= \frac{\Gamma(H^{\pm}\to  A^0W)}{ \Gamma(H^{\pm}\to  A^0W)+
\Gamma(H^{\pm}\to  tb)}
\label{R}
\end{eqnarray}
as a function of $\tan\beta$ for various values of $m_A$, fixing
$m_{H^\pm}=500$ GeV. We plot the tree-level width for 
$(H^{\pm}\to A^0W$) given in Eq.~2.3, 
and the tree-level width for $H^{\pm}\to  tb$ given in Ref.~\cite{djou3body},
assuming Model~II type couplings. Since other channels such as
$H^{\pm}\to h^0W$ and $H^\pm\to H^0W$ may be open the above ratio should be 
interpreted as the upper bound on BR$(H^{\pm}\to A^0W$). Note that
these additional channels are suppressed by the factors
$\cos^2(\beta-\alpha)$ and $\sin^2(\beta-\alpha)$ respectively, while
$\Gamma(H^{\pm}\to  A^0W)$ possesses no mixing angle suppression.
One can see that the decay $H^{\pm}\to A^0W$ is maximized for moderate
values of $\tan\beta$ (i.e. when $\Gamma(H^{\pm}\to tb$) is minimized).
The curves with lighter $m_A$ are less phase space suppressed and 
so the value of $R$ may be larger. For fixed $m_A$ and $\tan\beta$
the sensitivity to $m_{H^\pm}$ is rather mild. Larger $m_{H^\pm}$
slightly increases $R$ since $\Gamma(H^{\pm}\to  A^0W)\sim m_{H^\pm}^3$
while $\Gamma(H^{\pm}\to tb)\sim m_{H^\pm}$.

Given the possible large BR, an accurate measurement
of BR$(H^{\pm}\to  A^0W)$ may allow one to obtain information on 
$\lambda_5$. As explained above, the radiative corrections 
show sensitivity to several of the input parameters.  
If experimental information on the Higgs masses and mixing angles
were available then it might be possible to measure $\lambda_5$.
At a $e^+e^-$ linear collider \cite{lin} one could measure the
Higgs masses from a variety of production mechanisms 
($e^+e^-\to Zh^0,ZH^0,h^0A^0,H^0A^0,H^+H^-$). Information on the mixing angles
$\alpha,\beta$ could be obtained \cite{measure} from an analysis of
the production cross-sections and branching ratios. 
Other processes which are sensitive to $\lambda_5$ are
$e^+e^-\to H^+H^-$ \cite{ArhribMoult} and $H^{\pm}W^{\mp}$ \cite{HW},
while theoretical bounds on the Higgs masses in the case of $\lambda_5\ne 0$
are explored in Ref.~\cite{prepar},\cite{Okada}.

\renewcommand{\theequation}{6.\arabic{equation}}
\setcounter{equation}{0}
\section*{6. Conclusions}
We have computed the radiative corrections to the on-shell 
decay $H^+ \to A^0 W^+$ in the general Two Higgs Doublet Model,
taking into account the experimental constraint on the $\rho$ parameter 
and also unitarity constraints on the scalar sector parameters. We have
included the Yukawa corrections, the full electroweak corrections (bosonic), 
and also the real photon emission in the final state (Bremsstrahlung). 
The computation was done with dimensional regularization 
in the on-shell scheme. We find that the total radiative corrections
may approach 40\% in regions of parameter space for both small and 
moderate $\tan\beta$. The bosonic correction is 
sensitive to the soft discrete symmetry breaking parameter $\lambda_5$, 
and may interfere
both constructively and destructively with the Yukawa correction. For larger
$\lambda_5$ the bosonic contribution becomes strongly negative and
in general dominates the Yukawa correction. For $m_A\approx 2m_t$ and 
low $\tan\beta$ the Yukawa
correction is maximized and interferes constructively with the 
bosonic correction, resulting in large negative corrections 
to the tree-level width. 
Finally, we showed that the decay $H^{\pm}\to A^0W^{\pm}$ may supercede 
$H^{\pm}\to tb$ as the dominant decay channel, and thus a precise measurement
of its branching ratio may allow information to be obtained on $\lambda_5$.

\section*{Acknowledgements}
A.G.A was supported by the Japan Society for Promotion of Science (JSPS).
We thank C. Dove for reading the manuscript.

\input wha-all1.tex

\newpage
\renewcommand{\theequation}{C.\arabic{equation}}
\setcounter{equation}{0}

\subsection*{Figure Captions}

\vspace*{0.5cm}

\renewcommand{\labelenumi}{Fig. \arabic{enumi}}
\begin{enumerate}
\item  %Fig.1
Lowest-order Feynman diagram for the decay $H^+ \to A^0 W^+$.
\item  %Fig.2
Feynman diagrams for the one-loop corrections to the decay 
$H^+ \to A^0 W^{+}$: i) vertex (2.1 $\to $ 2.16), 
ii) Bremsstrahlung diagrams for  $H^+ \to A^0 W^+\gamma $: 
Fig. 2.17 $\to$ 2.19
 iii) CP-odd Higgs boson
 self-energy (2.20 $\to$ 2.28) and Charged
 Higgs boson self-energy (2.29 $\to$ 2.36).
\item  %Fig. 3
Total contribution to $\Gamma_H$ as function of $m_{H^\pm}$.
We chose: $m_H=m_{H^\pm}-10$, $m_h=120$, $m_A=150$ (GeV), 
$\alpha=\beta -\frac{\pi}{2}$.\\
Fig.3.a: $\tan\beta=0.5$, for four values of  $\lambda_5$=0.0, 2.0, 6.0
and 8.5.\\
Fig.3.b: $\tan\beta=1.5$, for four values of  
$\lambda_5$=0.0, 1.0, 3.0 and 5.0
 
\vspace{5mm}
\item  %Fig. 4
Total contribution to $\Gamma_H$ as function of $m_{A}$.
We chose: $m_H=500$, $m_h=360$, $m_{H^\pm}=530$ (GeV) and 
$\alpha=\beta -\frac{\pi}{2}$.\\
Fig.4.a: $\tan\beta=0.8$, for three values of  
$\lambda_5$=4.0, 6.0 
and 10.0.\\
Fig.4.b: $\tan\beta=1.6$, for three values of  
$\lambda_5$=4.0, 8.0 and 12.
   
\vspace{5mm}
\item  %Fig. 5  
\vspace{5mm}
Fig.5.a: Bosonic contribution ($\Gamma_H^{bos}$) to $\Gamma_H$ as 
function of $\lambda_5$.
We chose: $m_H=180$, $m_h=120$, $m_{H^\pm}=200$, $m_A=110$ (GeV), 
$\alpha=\beta -\frac{\pi}{2}$, and several values of $\tan\beta$.

Fig.5.b: Bosonic contribution ($\Gamma_H^{bos}$) to $\Gamma_H$ as 
function of $\sin\alpha$
for three different configurations $C_i$ i=1, 2, 3:\\
$C_1$ : $m_{H^\pm}=220$, $m_{H}=180$, $m_{h}=80$, $\tan\beta=3.6$ and $\lambda_5=5$\\
$C_2$ : $m_{H^\pm}=250$, $m_{H}=280$, $m_{h}=140$, $\tan\beta=1.6$ and $\lambda_5=5$\\
$C_3$ : $m_{H^\pm}=420$, $m_{H}=400$, $m_{h}=290$, $\tan\beta=2.6$ and $\lambda_5=8$

\vspace{5mm}
\item  %Fig. 6  
\vspace{5mm}
Ratio $R$ (eq \ref{R}) as a function of $\tan\beta$ for various values of
$m_{A}$ and for $m_{H^\pm}=500$ GeV.
\end{enumerate}

\newpage
\renewcommand{\thepage}{}
\begin{minipage}[t]{19.cm}
\setlength{\unitlength}{1.in}
\begin{picture}(0.1,0.1)(0.8,8.2)
\centerline{\epsffile{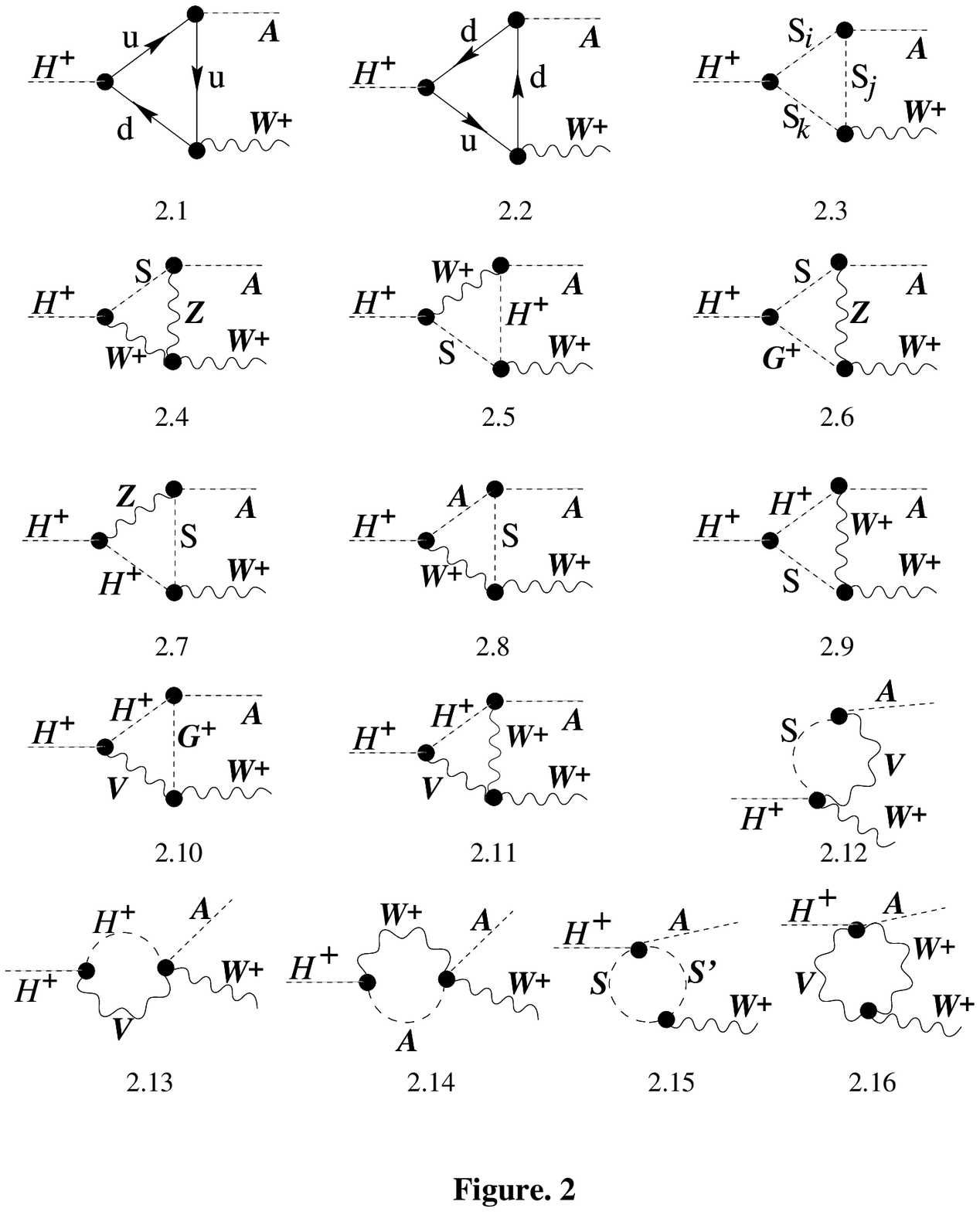}}
\end{picture}
\end{minipage}

%\newpage
%\begin{minipage}[t]{19.cm}
%\setlength{\unitlength}{1.in}
%\begin{picture}(0.1,0.1)(0.8,7.9)
%\centerline{\epsffile{gz.ps}}
%\end{picture}
%\end{minipage}
%\newpage
%\begin{minipage}[t]{19.cm}
%\setlength{\unitlength}{1.in}
%\begin{picture}(0.1,0.1)(0.8,7.9)
%\centerline{\epsffile{zz.ps}}
%\end{picture}
%\end{minipage}
%\newpage
%\begin{minipage}[t]{19.cm}
%\setlength{\unitlength}{1.in}
%\begin{picture}(0.1,0.1)(0.8,7.9)
%\centerline{\epsffile{ww.ps}}
%\end{picture}
%\end{minipage}

\newpage
\begin{minipage}[t]{19.cm}
\setlength{\unitlength}{1.in}
\begin{picture}(0.1,0.1)(0.8,7.9)
\centerline{\epsffile{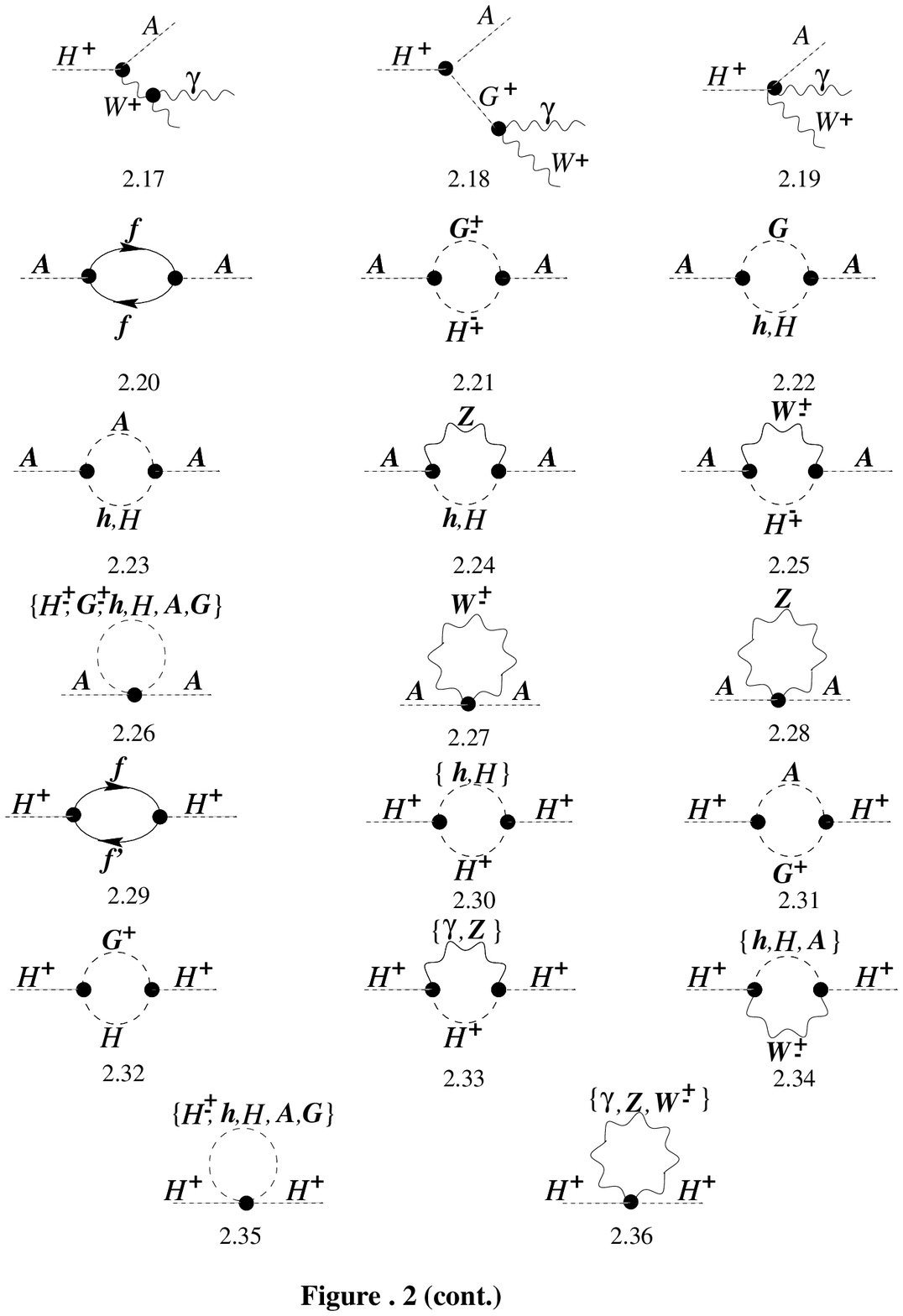}}
\end{picture}
\end{minipage}

\newpage
\begin{minipage}[t]{19.cm}
\setlength{\unitlength}{1.in}
\begin{picture}(1,1)(1.,9.1)
\centerline{\epsffile{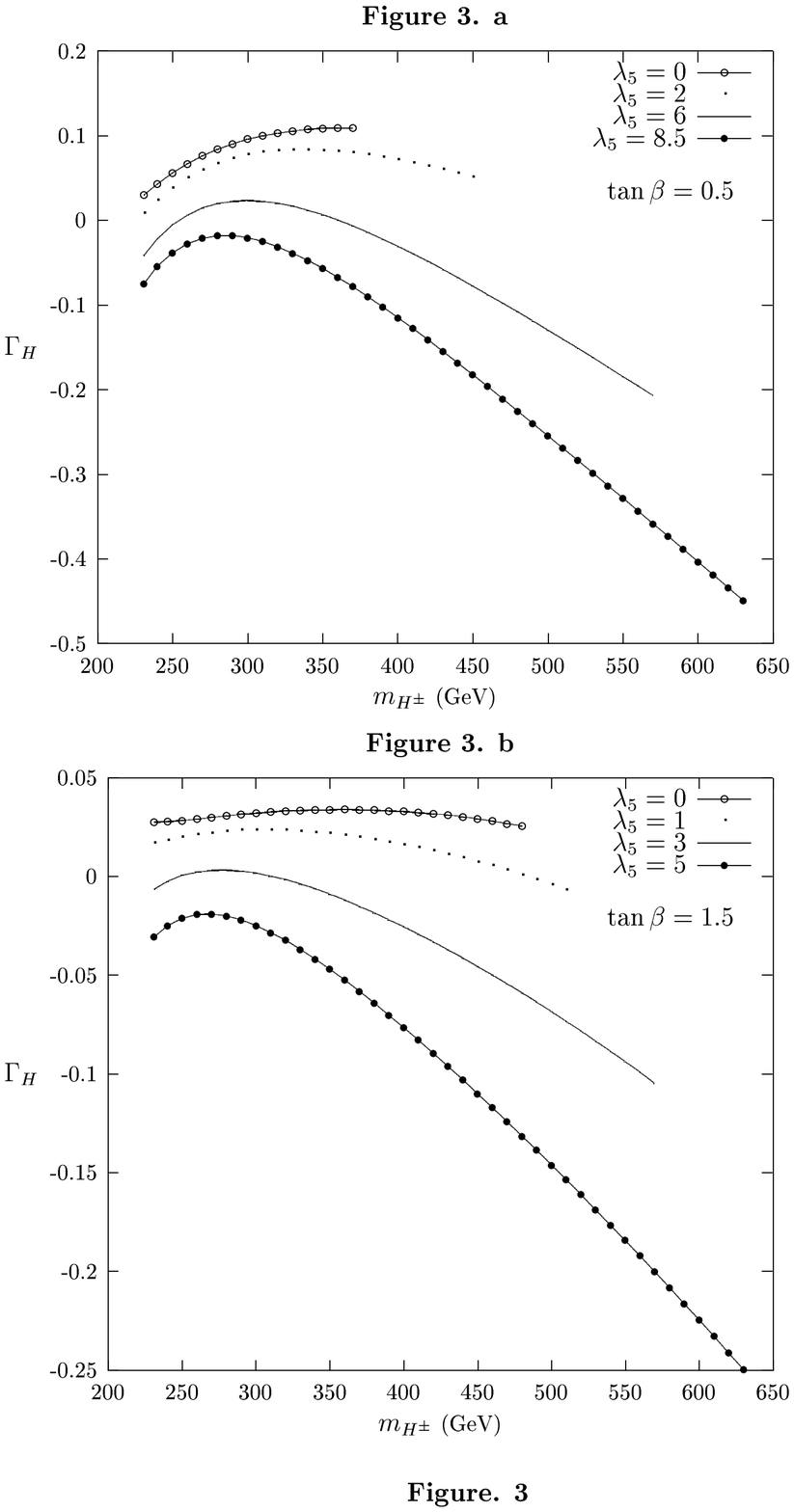}}
\end{picture}
\end{minipage}

\newpage
\begin{minipage}[t]{19.cm}
\setlength{\unitlength}{1.in}
\begin{picture}(1,1)(1.,9.1)
\centerline{\epsffile{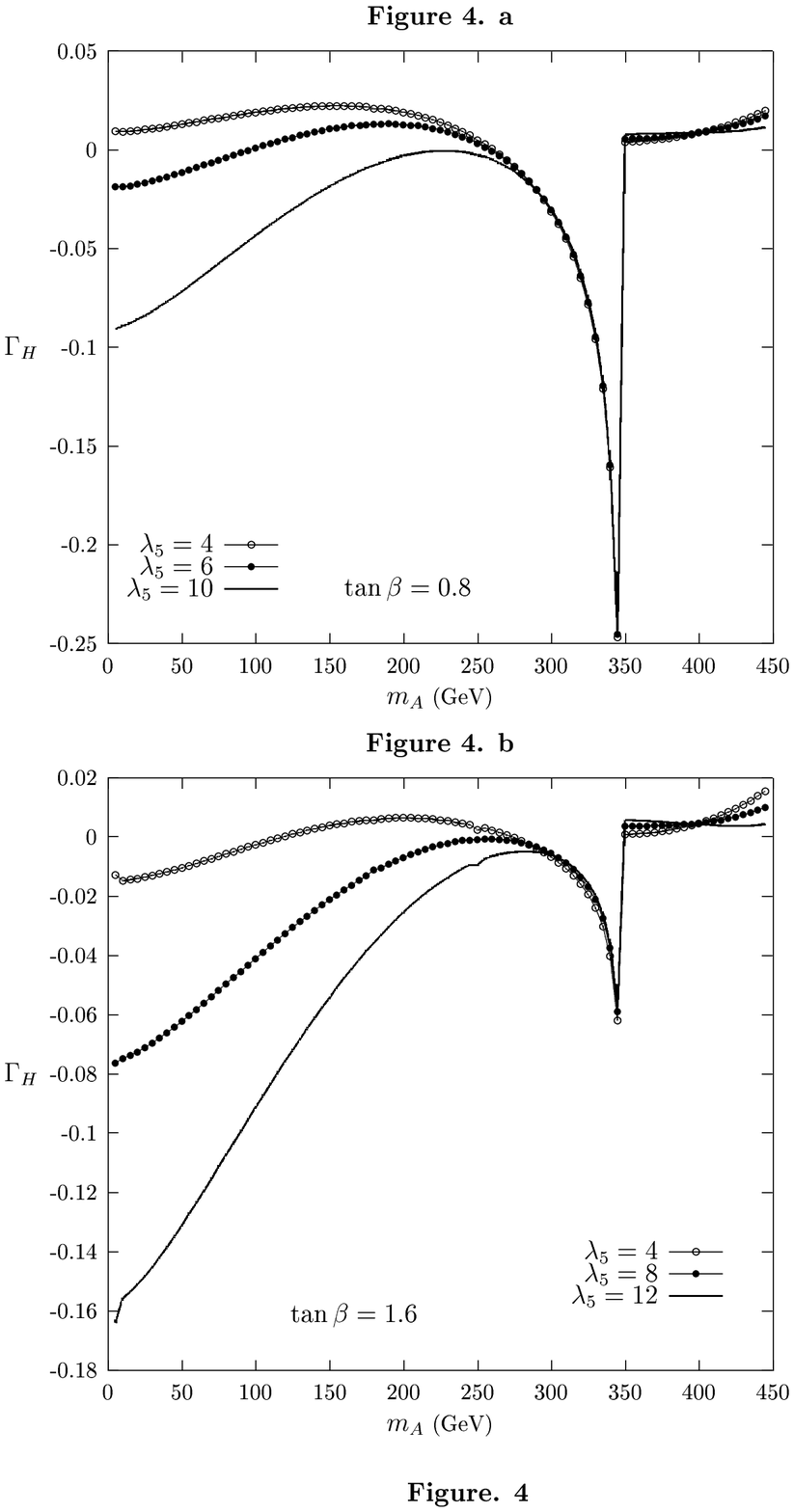}}
\end{picture}
\end{minipage}

\newpage
\begin{minipage}[t]{19.cm}
\setlength{\unitlength}{1.in}
\begin{picture}(1,1)(1.,9.1)
\centerline{\epsffile{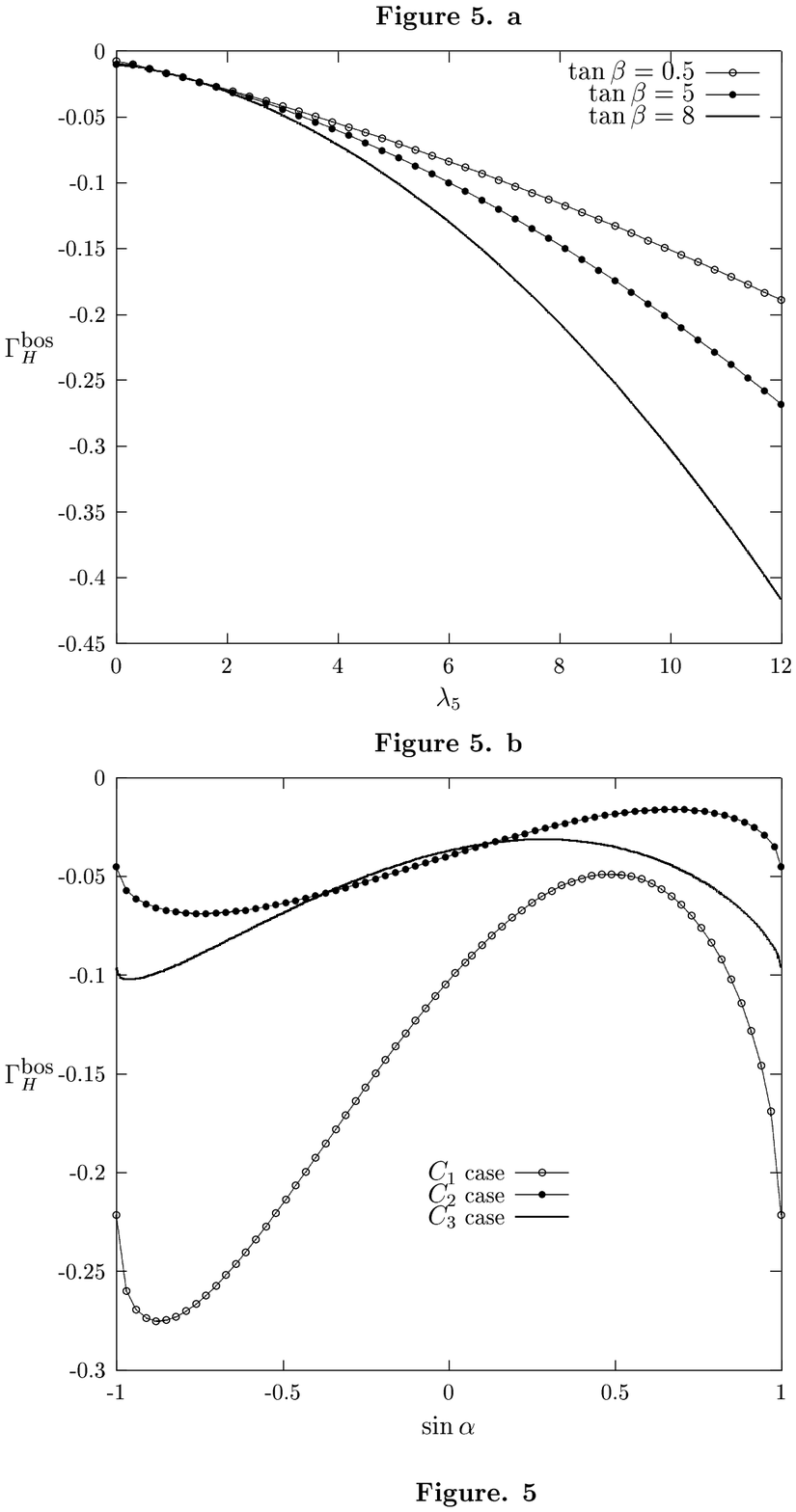}}
\end{picture}
\end{minipage}

\newpage
$\ $
\vspace{1.5cm}

\begin{minipage}[t]{19.cm}
\setlength{\unitlength}{1.in}
\begin{picture}(1,1)(1.,9.1)
\centerline{\epsffile{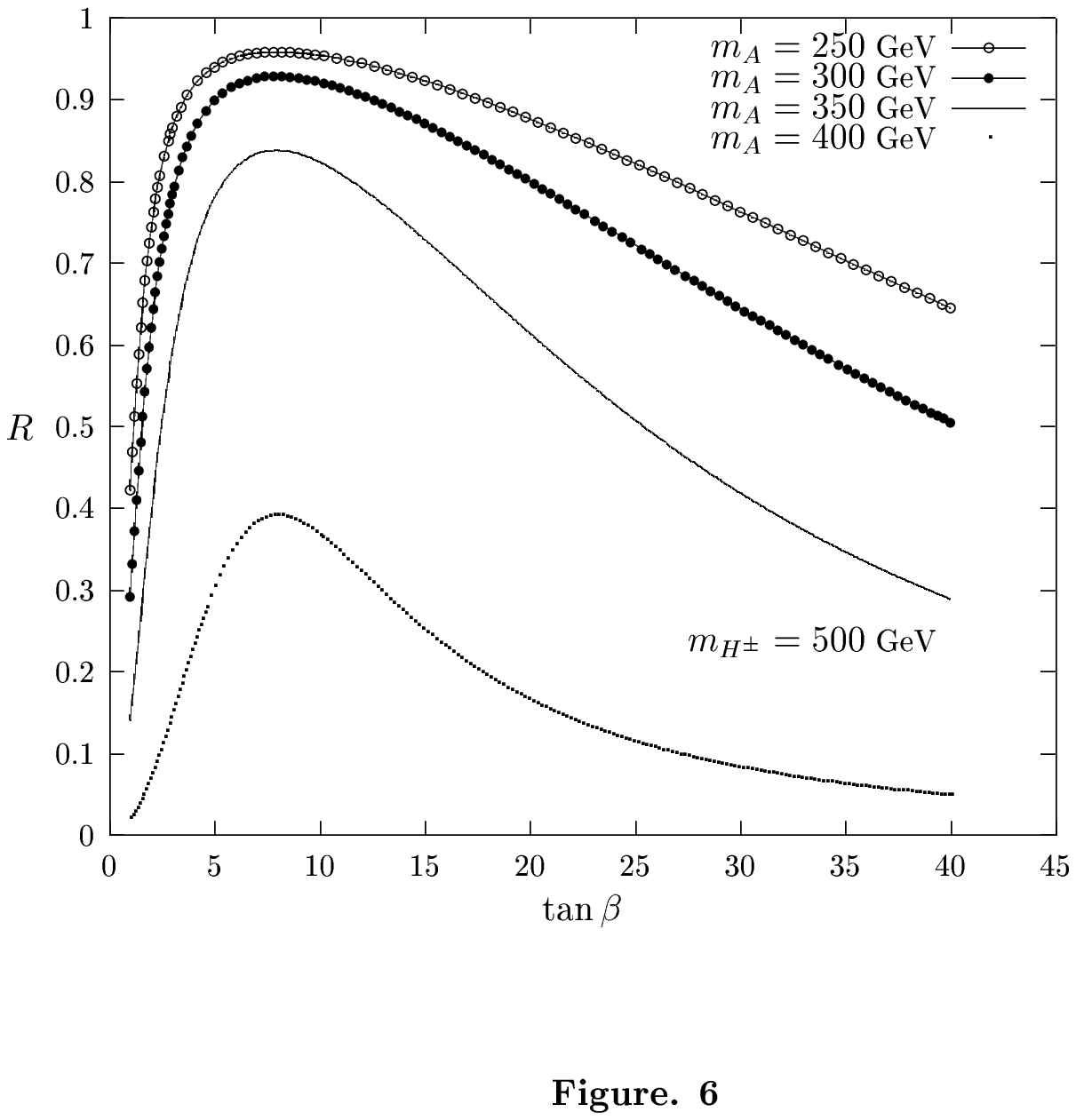}}
\end{picture}
\end{minipage}
$\ $
\vspace{5.cm}

\end{document}

%% file: fey1.tex
%%%%%% Begin   fey1
%Definition des diagrames de Feynman
% input of rpm.tex

%     un  Boson sortant:\unbs{}
\def\unbs#1{
\begin{picture}(100,200)(0,0)
\put(0,0){\circle*{5}}
\multiput(5,0)(8,-8){6}{\oval(8,8)[bl]}
\multiput(5,-8)(8,-8){6}{\oval(8,8)[tr]}
\put(49,-32){\makebox(0,0){#1}}
\end{picture}   }
%
%     Paire de  Bosons entrants:\be{}{}
\def\be#1#2{
\begin{picture}(100,200)(0,0)
\put(25,0){\circle*{5}}
\multiput(0,23)(8,-8){3}{\oval(8,8)[tr]}
\multiput(8,23)(8,-8){3}{\oval(8,8)[bl]}
\multiput(0,-23)(8,8){3}{\oval(8,8)[br]}
\multiput(8,-23)(8,8){3}{\oval(8,8)[tl]}
\put(-10,26){\makebox(0,0){#1}}
\put(-10,-26){\makebox(0,0){#2}}
\end{picture}   }
%   Paire de  Bosons sortants :\bs{}{}
\def\bs#1#2{
\begin{picture}(100,200)(0,0)
\put(0,0){\circle*{5}}
\multiput(5,0)(8,8){3}{\oval(8,8)[tl]}
\multiput(5,8)(8,8){3}{\oval(8,8)[br]}
\multiput(5,0)(8,-8){3}{\oval(8,8)[bl]}
\multiput(5,-8)(8,-8){3}{\oval(8,8)[tr]}
\put(34,30){\makebox(0,0){#1}}
\put(34,-30){\makebox(0,0){#2}}
\end{picture}   }
%  Paire de   Fermions entrants
\def\fe#1#2{
\begin{picture}(100,200)(0,0)
\put(34,0){\circle*{5}}
\put(0,35){\vector(1,-1){35}}
\put(-10,26){\makebox(0,0){#1}}
\put(-10,-26){\makebox(0,0){#2}}
\put(0,-35){\vector(1,1){35}}
\end{picture} }
%
%  Paire de   Fermions sortants
\def\fs#1#2{
\begin{picture}(100,200)(0,0)
\put(0,0){\circle*{5}}
\put(0,0){\vector(1,1){25}}
\put(34,26){\makebox(0,0){#1}}
\put(34,-26){\makebox(0,0){#2}}
\put(0,0){\vector(1,-1){25}}
\end{picture} }
% Boson propagateur :\bp{}{} propa:#2=24,entrant:#2=0,sortant#2=40
\def\bp#1#2{
\begin{picture}(100,200)(0,0)
\multiput(0,0)(16,0){3}{\oval(8,8)[t]}
\multiput(8,0)(16,0){3}{\oval(8,8)[b]}
\put(#2,15){\makebox(0,0){#1}}
\end{picture}   }
%
% Boson propagateur small :\sbp{}{} propa:#2=24,entrant:#2=0,sortant#2=40
\def\sbp#1#2{
\begin{picture}(100,200)(0,0)
\multiput(0,0)(16,0){2}{\oval(8,8)[t]}
\multiput(8,0)(16,0){2}{\oval(8,8)[b]}
\put(#2,15){\makebox(0,0){#1}}
\end{picture}   }
%
% Boson propagateur BIG :\bbp{}{} propa:#2=24,entrant:#2=0,sortant#2=40
\def\bbp#1#2{
\begin{picture}(100,200)(0,0)
\multiput(0,0)(16,0){5}{\oval(8,8)[t]}
\multiput(8,0)(16,0){5}{\oval(8,8)[b]}
\put(#2,15){\makebox(0,0){#1}}
\end{picture}   }
%
% Fermion propagateur :\fp{}{} propa:#2=24,entrant:#2=0,sortant#2=40
\def\fp#1#2{
\begin{picture}(100,200)(0,0)
\put(0,0){\line(1,0){48}}
\put(#2,15){\makebox(0,0){#1}}
\end{picture}   }
%
% Externel Fermion  :\fpe{}{} propa:#2=24,entrant:#2=0,sortant#2=40
\def\fpe#1#2{
\begin{picture}(100,200)(0,0)
\put(0,0){\line(1,0){35}}
\put(#2,15){\makebox(0,0){#1}}
\end{picture}   }
\def\fpp#1#2{
\begin{picture}(100,200)(0,0)
\put(0,0){\line(1,0){48}}
\put(#2,15){\makebox(0,0){#1}}
\end{picture}   }
%       Fermion voie t      :\ft{}
\def\ft#1{
\begin{picture}(100,200)(0,0)
\put(0,0){\line(0,1){48}}
\put(10,24){\makebox(0,0){#1}}
\end{picture}   }
\def\fth#1{
\begin{picture}(100,200)(0,0)
\put(0,0){\line(0,1){48}}
\put(10,24){\makebox(0,0){#1}}
\end{picture}   }
%       Boson voie t :\bt{}{}
\def\bt#1{
\begin{picture}(100,200)(0,0)
\multiput(0,0)(0,16){3}{\oval(8,8)[tl]}
\multiput(0,8)(0,16){3}{\oval(8,8)[br]}
\multiput(0,8)(0,16){3}{\oval(8,8)[tr]}
\multiput(0,0)(0,16){3}{\oval(8,8)[bl]}
\put(15,24){\makebox(0,0){#1}}
\end{picture}   }
\def\fpp#1#2{
\begin{picture}(100,200)(0,0)
\multiput(0,0)(10.8,0){6}{\line(1,0){5.2}}
\put(#2,15){\makebox(0,0){#1}}
\end{picture}   }
\def\hpe#1#2{
\begin{picture}(100,200)(0,0)
\multiput(0,0)(10.8,0){5}{\line(1,0){5.2}}
\put(#2,15){\makebox(0,0){#1}}
\end{picture}   }
\def\ftp#1{
\begin{picture}(100,200)(0,0)
\multiput(0,0)(0,10.8){5}{\line(0,1){5.2}}
\put(10,24){\makebox(0,0){#1}}
\end{picture}   }
%paire de bosons scalaires sortants
\def\fsp#1#2{
\begin{picture}(100,200)(0,0)
\put(0,0){\circle*{5}}
\multiput(0,0)(16,16){3}{\line(1,1){10}}
\multiput(0,0)(16,-16){3}{\line(1,-1){10}}
\put(50,36){\makebox(0,0){#1}}
\put(52,-36){\makebox(0,0){#2}}
\end{picture}   }
\def\fep#1#2{
\begin{picture}(100,200)(0,0)
\put(40,0){\circle*{5}}
\multiput(0,-40)(16,16){3}{\line(1,1){5}}
\multiput(0,40)(16,-16){3}{\line(1,-1){5}}
 \put(-7,36){\makebox(0,0){#1}}
\put(-7,-36){\makebox(0,0){#2}}
\end{picture}   }
%%%
% un boson scalaire sortants
\def\unfsp#1{
\begin{picture}(100,200)(0,0)
\put(0,0){\circle*{5}}
\multiput(0,0)(16,16){3}{\line(1,1){10}}
\put(45,36){\makebox(0,0){#1}}
\end{picture}   }
%%%%

%%%%%%%%%%%%%%%%%%%%%
%%%%%%%%%%%%%%%%%

%\end{document}

%% file: wha-all1.tex
%\documentstyle[12pt]{article}
%\oddsidemargin 0.26cm
%\evensidemargin 0.26cm
%\marginparwidth 68pt
%\marginparsep 10pt
%\topmargin 0cm
%\headheight 0pt
%\headsep 0pt
%\footskip 25pt
%\textheight 24cm
%\textwidth 16.cm
%\columnsep 10pt
%\columnseprule 0pt
%
%\begin{document}
\renewcommand{\theequation}{A.\arabic{equation}}
\setcounter{equation}{0}
\section*{Appendix A: THDM trilinear and quartic scalar couplings }
In this appendix we list the Feynman rules in the general THDM 
for the trilinear and quartic 
scalar couplings relevant for our study. All formulae are written in terms of 
the physical masses, $\alpha$, $\beta$ and the soft breaking term $\lambda_5$. 
Note that in the trilinear couplings (quartic couplings) we have factorised
out  $ie$ ($ie^2$). In the following
 $g_C=1/(2 s_W m_W s_{2\beta} )$ \ , 
\ $v^2 = \frac{2m_W^2}{g^2} $\ , \
$c_{\beta\alpha}^{\pm}=\cos (\beta{\pm}\alpha)$ and 
$s_{\beta\alpha}^{\pm}=\sin (\beta{\pm}\alpha)$. 
\subsubsection*{A.1 Trilinear scalar coupling}
\begin{eqnarray}
& & g_{H^0H^+H^-}=-2g_c( 
 m_{H^0}^2 (c_\beta^3 s_\alpha +s_\beta^3 c_\alpha)+m_{H^{\pm}}^2s_{2\beta}c_{\beta\alpha}^{-}-
s_{\beta\alpha}^+\lambda_5 v^2 )
\\
& & g_{H^0H^+G^-}  =  g_c s_{2\beta}s_{\beta\alpha}^- (m_{H^0}^2-m_{H^{\pm}}^2) \\
& & g_{h^0H^+H^-}  = - 2 g_c (m_{h^0}^2(c_{\alpha}c_{\beta}^3-
s_{\alpha}s_{\beta}^3) +m_{H^{\pm}}^2s_{2\beta}s_{\beta\alpha}^{-}-
c_{\beta\alpha}^+{\lambda_5} v^2)
\\
& & g_{h^0H^+G^-} =- g_c s_{2\beta} c_{\beta\alpha}^-(m_{h^0}^2-m_{H^{\pm}}^2)
\\
& & g_{H^0A^0A^0}  = - 2 g_c (m_{H^0}^2(s_{\alpha}c_{\beta}^3+c_{\alpha}s_{\beta}^3)+
m_A^2s_{2\beta}c_{\beta\alpha}^{-}-s_{\beta\alpha}^+\lambda_5 v^2 )
\\
& & g_{H^0A^0G^0} =  g_c s_{2\beta} s_{\beta\alpha}^-(m_{H^0}^2-m_A^2) \\
& & g_{h^0A^0A^0}  = - 2 g_c (m_{h^0}^2(c_{\alpha}c_{\beta}^3-s_{\alpha}s_{
\beta}^3)+m_A^2s_{2\beta}s_{\beta\alpha}^{-}-c_{\beta\alpha}^+\lambda_5 v^2)
\\
& & g_{h^0A^0G^0}  = g_c s_{2\beta} c_{\beta\alpha}^- (m_A^2-m_{h^0}^2)
\\
& & g_{A^0H^+G^-}  = i g_c s_{2\beta} (m_{H^{\pm}}^2 - m_A^2) 
\end{eqnarray}
\subsubsection*{A.2 Quartic scalar coupling}
\begin{eqnarray}
& & g_{H^0H^0H^0H^0}  = -12g_C^2 [m_{H^0}^2(c_{\beta}s_{\alpha}^3+s_{\beta}c_{\alpha}^3)^2+
m_{h^0}^2s_{\alpha}^2c_{\alpha}^2{s_{\beta\alpha}^-}^2-{\lambda_5}v^2(c_{\alpha}^2-c_{\beta}^2)^2]
\\ & & 
g_{h^0h^0h^0h^0}  =  -12 g_C^2  [m_{H^0}^2s_{\alpha}^2c_{\alpha}^2{c_{\beta\alpha}^-}^2
+m_{h^0}^2(c_{\beta}c_{\alpha}^3-s_{\beta}s_{\alpha}^3)^2-{\lambda_5}{v}^2(c_{\alpha}^2-s_{\beta}^2)^2 ]
\\ & & 
g_{A^0A^0A^0A^0}  = -12 g_C^2 [m_{H^0}^2(c_{\alpha}s_{\beta}^3 + s_{\alpha}c_{\beta}^3)^2 + 
m_{h^0}^2(c_{\alpha}c_{\beta}^3-s_{\alpha}s_{\beta}^3)^2-{\lambda_5}v^2c_{2\beta}^2 ]
\\ & & 
g_{G^0G^0G^0G^0}  =  - 3 g_C^2 s_{2\beta}^2 [
m_{H^0}^2{c_{\beta\alpha}^-}^2+m_{h^0}^2{s_{\beta\alpha}^-}^2 ]
\\ & & 
g_{H^+H^-H^+H^-}  =  -8 g_C^2 [m_{H^0}^2(c_{\alpha}s_{\beta}^3 + s_{\alpha}c_{\beta}^3)^2 + 
m_{h^0}^2(c_{\alpha}c_{\beta}^3-s_{\alpha}s_{\beta}^3)^2-\lambda_5v^2c_{2\beta}^2 ]
\\ & & 
g_{G^+G^-G^+G^-}  =  -2 g_C^2  s_{2\beta}^2  [m_{H^0}^2{c_{\beta\alpha}^-}^2+
m_{h^0}^2{s_{\beta\alpha}^-}^2 ] \\ & &
g_{H^0H^0h^0h^0}  =  -4 g_C^2  [m_{H^0}^2 s_{\alpha}c_{\alpha} (s_\beta c_\beta + 
3s_{\alpha}c_{\alpha} {s_{\beta\alpha}^-}^2)-m_{h^0}^2 s_{\alpha}c_{\alpha}
( s_\beta c_\beta - 3s_{\alpha}c_{\alpha}  {c_{\beta\alpha}^-}^2)\nonumber \\ & & 
\qquad \qquad  -\lambda_5v^2
(3s_{\alpha}^2c_{\alpha}^2-s_{\beta}^2c_{\beta}^2)]\\ &&
g_{H^0H^0A^0A^0}=-2g_C^2[2m_{H^0}^2(s_{\alpha}c_{\beta}^3+
c_{\alpha}s_{\beta}^3)(c_{\beta}s_{\alpha}^3+s_{\beta}c_{\alpha}^3)+
2m_{h^0}^2s_{\alpha}c_{\alpha}s_{\beta\alpha}^{-}(c_{\alpha}c_{\beta}^3-s_{\alpha}s_{\beta}^3)
\nonumber \\& &  \qquad \qquad 
+m_A^2(c_{\beta}^2-s_{\beta}^2)(c_{\alpha}^2-s_{\beta}^2+2s_{\alpha}s_{\beta}{c_{\beta\alpha}^-} ) 
\nonumber \\ & &  \qquad \qquad +2\lambda_5v^2(c_{2\beta}(c_{\alpha}^2s_{\beta}^4-
s_{\alpha}^2c_{\beta}^4)-s_{\beta}^2c_{\alpha}^2(1+s_{2\alpha}s_{2\beta}))] \\ & &
g_{h^0h^0A^0A^0} = -2g_C^2 [2m_{H^0}^2(s_{\alpha}c_{\beta}^3+c_{\alpha}s_{\beta}^3)
s_{\alpha}c_{\alpha}c_{\beta\alpha}^{-}+2m_{h^0}^2(c_{\beta}c_{\alpha}^3-s_{\beta}
s_{\alpha}^3)(c_{\alpha}c_{\beta}^3-s_{\alpha}s_{\beta}^3) \nonumber \\ & & \qquad \qquad 
+m_A^2(c_{\beta}^2-s_{\beta}^2)(s_{\alpha}^2+s_{\beta}^2-2s_{\alpha}s_{\beta}{c_{\beta\alpha}^-})
\nonumber \\  & & \qquad \qquad
+2\lambda_5v^2(c_{2\beta}(s_{\alpha}^2s_{\beta}^4-c_{\alpha}^2c_{\beta}^4)-
s_{\beta}^2c_{\alpha}^2(1-s_{2\alpha}s_{2\beta})) ] 
\end{eqnarray}

\renewcommand{\theequation}{B.\arabic{equation}}
\setcounter{equation}{0}
\section*{Appendix B: One-loop vertex: $H^+\to W A^0$}
In this appendix we will list the analytic expression
for each generic diagram of Fig.2. The sum over all the 
particle contents yields the corresponding 
contribution to the one-loop amplitude for 
$H^+\to W A^0$. For scalar and tensor integrals we use
the same convention as \cite{ArhribMoult}, the analytical 
expression of all scalar functions can be found in \cite{passarino,ff}.

\subsection*{B. 1 Fermionic Loops}
The diagram with the $uud$ triangle, Fig.2.1, yields the following
contribution to the one-loop amplitude:
 \begin{eqnarray}
M_{2.1} & = & -\frac{e\alpha N_C}{2\sqrt{2}\pi s_W}Y_{uu}
      \{ Y_{ud}^L ( 2 B_0 +  B_1  )
+ m_u C_0(m_u Y_{ud}^L+ m_d Y_{ud}^R ) \nonumber\\ &&
+ Y_{ud}^L[m_W^2 C_2 - 2 C_{00} 
- 2 m_A^2 C_{11} 
+ C_{12}(m_{H^{\pm}}^2- m_W^2- m_A^2)]\}
\end{eqnarray}
$N_C=3(1)$ for quarks (leptons).
Therein, all the $B_0$, $B_1$, $C_i$ and $C_{ij}$ have the same arguments:
$$[ B_0 ,B_1](m_{H^{\pm}}^2,m_u^2,m_d^2)\ , \ 
[C_i , C_{ij}](m_{A}^2,m_{H^{\pm}}^2,p_{W}^2,m_u^2,m_u^2,m_d^2)$$
The summation has to be performed over all fermion families; in 
practice only the third quark generation is relevant. 
\\
\\
 The corresponding expression for the 
the diagram with the $ddu$ triangle in Fig.2.2 
is obtained from the previous one by making the following replacements:
$$ Y_{ud}^L \longleftrightarrow Y_{ud}^R \ \ , \ \ Y_{uu}\longleftrightarrow Y_{dd} 
\ \ , \ \ m_u \longleftrightarrow m_d $$

\subsection*{B. 2 Bosonic Loops}
\subsection*{Diagram 2.3}
\begin{eqnarray}
M_{2.3} &=& \frac{e\alpha}{2\pi }g_{H^+S_iS_k}g_{AS_i S_j}g_{W^+S_j S_k }  
C_1(m_A^2,m_{H^\pm}^2,m_W^2,m_{S_j}^2,m_{S_i}^2,m_{S_k}^2)
\end{eqnarray}
The couplings are summarized in the following table:\\
\begin{center}
\begin{tabular}[t]{|c|c|c|c|} \hline 
$( S_i,S_j,S_k)$   &     $g_{H^+S_{i}S_k}$ &     $g_{AS_iS_j}$ & $g_{W^+S_jS_k}$ \\ \hline
$(h,A,H^+)$      & $g_{hH^+H^+}$    & $g_{hAA}$    & $\frac{1}{2s_W}$ \\ \hline
$(G^+,H^+,h)$    & $g_{hH^+G^+}$        & $g_{AG^+H^+}$      & $\frac{c_{\beta\alpha}^-}{2s_W}$ \\ \hline
$(G^+,H^+,H)$     & $g_{HH^+G^+}$        & $-ig_{AH^+G^+}$     & $-\frac{s_{\beta\alpha}^-}{2s_W}$ \\ \hline
$(G^+,H^+,A)$     & $-ig_{AH^+G^+}$        & $-ig_{AH^+G^+}$  & $\frac{1}{2s_W}$  \\ \hline
$(H,A,H^+)$     &  $g_{HH^+H^+}$        & $g_{HAA}$     & $\frac{1}{2s_W}$ \\  \hline
$(h,G^0,G^+)$     & $g_{hH^+G^+}$        & $g_{hAG^0}$     & $\frac{1}{2s_W}$ \\ \hline
$(H,G^0,G^+)$     & $g_{HH^+G^+}$        & $g_{HAG^0}$     & $\frac{1}{2s_W}$ \\ \hline
$(A,h,G^+)$     & $-ig_{AH^+G^+}$        & $g_{hAA}$      &  $\frac{s_{\beta\alpha}^-}{2s_W}$ \\ \hline
$(A,H,G^+)$     & $-ig_{AH^+G^+}$        & $g_{HAA}$      & $\frac{c_{\beta\alpha}^-}{2s_W}$ \\ \hline
$(H^+,G^+,h)$     & $ g_{hH^+H^+}$        & $-ig_{AH^+G^+}$     & $-\frac{s_{\beta\alpha}^-}{2s_W}$ \\ \hline
$(H^+,G^+,H)$     & $g_{HH^+H^+}$        & $-ig_{AH^+G^+}$     & $ -\frac{c_{\beta\alpha}^-}{2s_W}$ \\ \hline
\hline
\end{tabular}
\end{center}
\subsection*{Diagram 2.4}
\begin{eqnarray}
 M_{2.4}[h] & = &  -\frac{e{\alpha}{c_{\beta\alpha}^-}^2}{16{\pi}s_W^3 }
\{ -4B_0(m_{H^\pm}^2,m_h^2,m_W^2)
+[2(- m_A^{2} + m_{H^\pm}^{2} - 2m_Z^2) - 4p_W^{2}] C_{0}[h]
\nonumber\\ & &
 +[ 3(- m_A^{2} + m_{H^\pm}^{2})- 4p_W^{2}] C_{1}[h]
+[4( - m_A^{2} + m_{H^\pm}^{2})- 4p_W^{2}]  C_{2}[h]+ 4C_{00}[h]
\nonumber\\ &&
  + [m_{H^\pm}^{2} +  
3m_A^{2}-p_W^{2} ]  C_{11}[h]
-2[ m_{H^\pm}^{2}- m_A^{2}]  C_{12}[h]\} \\
M_{2.4}[H] & = & M_{2.4}[h] [{c_{\beta\alpha}^-}^2 \to 
{s_{\beta\alpha}^-}^2 ,m_h \to m_H , h \to H ]
\end{eqnarray}
Where the arguments of $C_i$ and $C_{ij}$ are given as follows:\\
$$[C_i,C_{ij}][S] =[C_i,C_{ij}](m_A^2,m_{H^\pm}^2,p_W^2,m_Z^2,m_S^2,
m_{W}^2)$$
\subsection*{Diagram.2.5}
\begin{eqnarray}
 M_{2.5}[h] & = & -\frac{e{\alpha}{c_{\beta\alpha}^-}^2}{16{\pi}s_W^3}
\{ B_0(m_{H^\pm}^{2},m_W^{2},m_h^{2}) + B_1(m_{H^\pm}^{2},m_W^{2},m_h^{2}) + 
[ p_W^{2} - 2 m_{H^\pm}^{2} ] C_{1}[h] \nonumber\\ &&
+ 2C_{00}[h]+ [ m_{H^\pm}^{2}  - (p_W^{2} - m_A^{2} )] C_{11}[h]
- [p_W^{2} - m_A^{2} + m_{H^\pm}^{2}]  C_{12}[h]\}
\\
 M_{2.5}[H] & = & M_{2.5}[h][ c_{\beta\alpha}^- \to s_{\beta\alpha}^- ,m_h \to m_H , h \to H ]
\\
M_{2.5}[A] & = & M_{2.5}[h][{c_{\beta\alpha}^-}^2 \to -1 ,m_h \to m_A , h \to A ]
\end{eqnarray}
Where the arguments of $C_i$ and $C_{ij}$ 
are as follows:
$$[C_i,C_{ij}][S]=[C_i,C_{ij}](m_A^2,m_{H^\pm}^2,p_W^2,m_{H^\pm}^2,m_W^2,
m_S^2)$$
\subsection*{Diagram.2.6}
\begin{eqnarray}
  M_{2.6}[h] & = & \frac{e{\alpha}c_{\beta{\alpha}^-}m_W}{8{\pi}c_W^2}
g_{h^0H^+G^-}(2C_{0}[h] +C_{1}[h])
\\
M_{2.6}[H] & = &  M_{2.6}[h][c_{\beta\alpha}^- \to -s_{\beta\alpha}^- , h \to H ]
\end{eqnarray}
Where $C_0$ and $C_1$ have the same arguments as follows: \\
$$ [C_0\ ,\ C_1][S]=[C_0\ ,\ C_1](m_A^2,m_{H^\pm}^2,p_W^2,m_Z^2,m_S^2,m_{W}^2)$$.
\subsection*{Diagram.2.7}
\begin{eqnarray} 
M_{2.7}[h] & = & -\frac{e{\alpha}(c_W^2-s_W^2) {c_{\beta\alpha}^-}^2}{16{\pi}s_W^3c_W^2}\{ 
B_0(m_{H^\pm}^{2},m_Z^{2},m_{H^\pm}^{2}) +  
B_1(m_{H^\pm}^{2},m_Z^{2}, m_{H^\pm}^{2} ) \nonumber\\ & & 
+[p_W^{2}-m_{h}^{2} -m_{H^\pm}^{2}] C_{1}[h] 
+2C_{00}[h] + [-p_W^{2} + m_A^{2} + m_{H^\pm}^{2}]  C_{11}[h] \nonumber\\ & &
-  [p_W^{2} - m_A^{2} + m_{H^\pm}^{2}]   C_{12}[h] \} \\
M_{2.7}[H] & = & M_{2.7}[h] [c_{\beta\alpha}^- \to s_{\beta\alpha}^- , 
h \to H , m_h\to m_H ]
\end{eqnarray}
The arguments of $C_i$ and $C_{ij}$ are given by :
$$[C_i,C_{ij}][S]=[C_i,C_{ij}](m_A^2,m_{H^\pm}^2,p_W^2,m_S^2,m_Z^2,m_{H^\pm}^2)$$
\subsection*{Diagram.2.8}
\begin{eqnarray} 
 M_{2.8}[h] & = & \frac{e{\alpha}s_{\beta{\alpha}^-}m_W}{8{\pi}s_W^2}g_{h^0A^0A^0}
(2C_{0}[h] +  C_{1}[h])
\\
 M_{2.8}[H] & = & M_{2.8}[h] [s_{\beta\alpha}^- \to c_{\beta\alpha}^- , 
h \to H ,g_{h^0A^0A^0}\to g_{H^0A^0A^0}  ]
\end{eqnarray}
Where $C_0$ and $C_1$ have the same arguments:\\
$$[C_0,C_1][S] = [C_0,C_1](m_A^2,m_{H^\pm}^2,p_W^2,m_S^2,m_A^2,m_{W}^2)$$.
\subsection*{Diagram.2.9}
\begin{eqnarray} 
 M_{2.9}[h] & = & \frac{e{\alpha}s_{\beta\alpha}^- m_W}{8{\pi}s_W^2}g_{h^0H^+H^-}
(2C_{0}[h] +  C_{1}[h])
\\
 M_{2.9}[H] & = & M_{2.9}[h] [s_{\beta\alpha}^- \to c_{\beta\alpha}^- , 
h \to H, g_{h^0H^+H^-}\to g_{H^0H^+H^-} ]
\end{eqnarray}
Where $C_0$ and $C_1$ have the same arguments:\\
$$[C_0,C_1][S] = [C_0,C_1](m_A^2,m_{H^\pm}^2,p_W^2,m_W^2,m_{H^\pm}^2,m_S^2)$$.
\subsection*{Diagram.2.10}
\begin{eqnarray}
 M_{2.10}[Z] & = & -\frac{e{\alpha}(c_W^2-s_W^2)m_W}{8{\pi}c_W^2}g_{A^0H^+G^-}
(2C_{0}[Z] +  C_{1}[Z])
\\
M_{2.10}[\gamma] & = & \frac{e{\alpha}  m_W}{4{\pi}}g_{A^0H^+G^-}
(2C_{0}[\delta ] +  C_{1}[\delta])
\end{eqnarray}
Again the $C_0$ and $C_1$ have the same arguments: \\
 $$[ C_0\ , \ C_1 ][V] = [ C_0\ , \ C_1 ](m_A^2,m_{H^\pm}^2,p_W^2,m_W^2,
m_{H^\pm}^2,m_V^2)$$
Here $\delta$ is a small photon mass introduced to regularise the
infrared divergence contained in the $C_0$ function. 
\subsection*{Diagram.2.11}
\begin{eqnarray} 
  M_{2.11}[Z] & = & \frac{e{\alpha}(c_W^2-s_W^2)}{16{\pi}s_W^3}
( 4B_0(m_{H^\pm}^2,m_Z^2,m_{H^\pm}^2)
 +  [ 2p_W^{2} +m_A^{2} - m_{H^\pm}^{2} +  2m_W^{2} ]  2C_{0}[Z]
\nonumber\\ &&
+ [4p_W^{2} + 3( m_A^{2} - m_{H^\pm}^{2}) ]  C_{1}[Z]
+  [p_W^{2} + m_A^{2} - m_{H^\pm}^{2}) ]  4C_{2}[Z]- 4C_{00}[Z]
\nonumber\\ &&
 +[p_W^{2} -  m_{H^\pm}^{2}  -  3 m_A^{2}  ]  C_{11}[Z]
+[ m_{H^\pm}^{2} - m_A^2  ] 2C_{12}[Z])
\\
  M_{2.11}[\gamma] & = & M_{2.11}[Z][\frac{(c_W^2-s_W^2)}{s_W^2} 
\to 2 ,m_Z \to \delta ]
\end{eqnarray}
All $C_i$ and $C_{ij}$ have the same arguments
$$[ C_i\ , \ C_{ij} ][V] = [ C_i\ , \ C_{ij} ](m_A^2,m_{H^\pm}^2,p_W^2,m_W^2,
m_{H^\pm}^2,m_V^2)$$
\subsection*{Diagram.2.12}
\begin{eqnarray}
M_{2.12}[h,Z] & = & -\frac{e{\alpha}{c_{\beta\alpha}^-}^2}{16{\pi}s_Wc_W^2} 
(2B_0(m_A^{2},m_Z^{2},m_h^{2})+B_1(m_A^{2},m_Z^{2},m_h^{2}))  \\
M_{2.12}[H^+,W^+] & = & -\frac{e{\alpha}}{16{\pi}s_W^3} (
2B_0(m_A^{2},m_W^{2},m_{H^\pm}^{2})+B_1(m_A^{2},m_W^{2},m_{H^\pm}^{2}))  \\
M_{2.12}[H,Z] & = & -\frac{e{\alpha}{s_{\beta\alpha}^-}^2}{16{\pi}s_Wc_W^2} (
2B_0(m_A^{2},m_Z^{2},m_H^{2})+B_1(m_A^{2},m_Z^{2},m_H^{2})) 
\end{eqnarray}
\subsection*{Diagram.2.13}
\begin{eqnarray}
 & & M_{2.13}[Z]  =  \frac{e{\alpha}(c_W^2-s_W^2)}{16{\pi}s_Wc_W^2}(B_0
(m_{H^\pm}^{2},m_{H^\pm}^2,m_Z^2)- 
B_1(m_{H^\pm}^{2},m_{H^\pm}^2,m_Z^2)) 
\\
& & M_{2.13}[\gamma]  =  -\frac{e{\alpha}}{8{\pi}s_W}(B_0(m_{H^\pm}^{2},
m_{H^\pm}^2,\delta)- 
B_1(m_{H^\pm}^{2},m_{H^\pm}^2,\delta))
\end{eqnarray}
\subsection*{Diagram.2.14 }
\begin{eqnarray}
 M_{2.14} &= & -\frac{e\alpha}{16{\pi}s_W^3}(2B_0(m_{H^{\pm}}^2,m_W^2,m_A^2)+ 
B_1(m_{H^{\pm}}^2,m_W^2,m_A^2)) 
\end{eqnarray}
\subsection*{Diagram 2.15 and 2.16}
For this kind of topology, it is clear that the 
amplitude is proportional to the W gauge boson momentum 
(Lorentz invariance) and consequently for the W on-shell 
the amplitude vanishes.
\begin{eqnarray}
 M_{2.15} =  0 \qquad, \qquad M_{2.16}=0 
\end{eqnarray}

\renewcommand{\theequation}{C.\arabic{equation}}
\setcounter{equation}{0}
\section*{Appendix C: Charged and CP-odd Higgs bosons self-energies}
This appendix is devoted to the self-energies of the 
charged and CP-odd Higgs bosons which are needed for the on-shell 
renormalisation scheme. The gauge bosons self energies
$\gamma$-$\gamma$, $\gamma$-$Z$, $Z$-$Z$ and $W$-$W$ can be found 
in \cite{gauge}.

\subsection*{C.1 CP-odd Higgs boson self-energy}
The CP-odd Higgs self-energy $\Sigma^{AA}$ 
can be cast into three parts: i) fermionic part 2.20, ii) pure scalar part
2.21; 2.22, 2.23 and 2.26  iii) 
mixing of gauge boson and scalar 2.24, 2.25 , 2.27 and 2.28 in such 
a way that:
\begin{eqnarray}
& & \Sigma^{AA}(q^2)=\Sigma_f^{AA}(q^2)+ \Sigma_S^{AA}(q^2)+
\Sigma_{VS}^{AA}(q^2)
\end{eqnarray}
with
\begin{eqnarray}
& & \Sigma_{f}^{AA}(q^2) = -\frac{{\alpha}N_C}{\pi}Y_{ff}^2(A_0(m_f^2) + 
 q^2 B_1(q^2, m_f^2, m_f^2)) \\ 
& & \Sigma_{S}^{AA}(q^2) =\frac{\alpha}{4\pi}(g_{hAA}^2B_0(q^2, m_A^2, m_h^2) +  
g_{HAA}^2B_0(q^2, m_A^2, m_H^2)\nonumber\\ && \qquad \qquad + 
 g_{hAG}^2B_0(q^2, m_h^2, m_Z^2) + g_{HAG}^2B_0(q^2, m_H^2, m_Z^2) \nonumber\\ & &  
\qquad \qquad + 
2g_{AH^+G^-}^2B_0(q^2, m_{H^{\pm}}^2, m_W^2))
+ \frac{\alpha}{8{\pi} } ( - g_{AAAA}A_0(m_A^2)
- g_{hhAA}A_0(m_h^2)\nonumber\\ && \qquad\qquad -  g_{HHAA}A_0(m_H^2) 
-2 g_{H^+H^-AA}A_0(m_{H^{\pm}}^2)    
-2 g_{G^+G^-AA}A_0(m_W^2) \\ && \qquad\qquad - g_{AAGG}A_0(m_Z^2)
+ \frac{2}{s_W^2c_W^2}A_0(m_Z^2)+ \frac{4}{s_W^2}A_0(m_W^2)
 -\frac{2}{s_W^2}m_W^2 - \frac{1}{s_W^2c_W^2}m_Z^2 )\nonumber 
\\ 
& & \Sigma_{VS}^{AA}(q^2) =  - \frac{\alpha}{16{\pi}c_W^2s_W^2}({c_{\beta\alpha}^-}^2
(A_0(m_Z^2) + (m_h^2 +q^2) B_0(q^2, m_h^2, m_Z^2) 
\nonumber\\ && \qquad\qquad
 -2q^2B_1(q^2, m_h^2, m_Z^2))+ 
 {s_{\beta\alpha}^-}^2(A_0(m_Z^2) + 
(m_H^2 +q^2) B_0(q^2, m_H^2, m_Z^2) \nonumber\\ && 
\qquad\qquad
-2q^2B_1(q^2, m_H^2, m_Z^2)) + c_W^2(A_0(m_W^2)  + 
(m_{H^{\pm}}^2 + q^2)B_0(q^2, m_{H^{\pm}}^2, m_W^2) \nonumber\\ && 
\qquad\qquad
-2q^2B_1(q^2, m_{H^{\pm}}^2, m_W^2)))
\end{eqnarray}

\subsection*{C.2 Charged Higgs boson self-energy}
The charged Higgs boson self-energy $\Sigma^{H^+H^-}$ 
can be cast into three parts: i) fermionic part 2.29, ii) 
pure scalar part 2.30; 2.31, 2.32 and 2.35
iii) mixing of gauge boson and scalar 2.33, 2.34 
and 2.36 in such a way that:
\begin{eqnarray}
& & \Sigma^{H^+H^-}(q^2)=\Sigma_f^{H^\pm H^\pm}(q^2)+ \Sigma_S^{H^+H^-}(q^2)+
\Sigma_{VS}^{H^+H^-}(q^2)
\end{eqnarray}
\begin{eqnarray}
& & \Sigma_{ff^{\prime}}^{H^{\pm}H^{\pm}}(q^2)  =   
-\frac{{\alpha}N_C}{2\pi}( ({Y_{ff^{\prime}}^L}^2+ {Y_{ff^{\prime}}^R}^2)(A_0(m_f^2) + q^2 B_1
(q^2, m_{f^{\prime}}^2, m_f^2) )
\nonumber\\ && \qquad\qquad \ \
+ (m_{f^{\prime}}^2 ( {Y_{ff^{\prime}}^L}^2+ 
{Y_{ff^{\prime}}^R}^2 )+
 2 m_{f^{\prime}}m_f Y_{ff^{\prime}}^L Y_{ff^{\prime}}^R)B_0
(q^2, m_{f^{\prime}}^2, m_f^2) )
\\  
& & \Sigma_{S}^{H^{\pm}H^{\pm}}(q^2)  = \frac{\alpha}{4\pi}(g_{AH^+G^-}^2B_0(q^2, m_A^2, m_W^2) + 
       g_{hH^+H^-}^2B_0(q^2, m_h^2, m_{H^{\pm}}^2) 
\nonumber\\ && \qquad\qquad\ \
+ g_{hH^+G^-}^2B_0(q^2, m_h^2, m_W^2) 
 + g_{HH^+H^-}^2B_0(q^2, m_H^2, m_{H^{\pm}}^2)
\nonumber\\ && \qquad\qquad\ \
+ g_{HH^+G^-}^2B_0(q^2, m_H^2, m_W^2)) -\frac{\alpha}{8\pi}(g_{H^+H^-AA}A_0(m_A^2) 
+g_{H^+H^-hh}A_0(m_h^2) \nonumber\\ &&
\qquad\qquad\ \ +g_{H^+H^-HH}A_0(m_H^2) 
+2g_{H^+H^-H^+H^-}A_0(m_{H^{\pm}}^2)
+\frac{2}{s_W^2}(m_W^2 - 2A_0(m_W^2)) \nonumber\\ &&
\qquad\qquad \ \ 
+g_{H^+H^-GG}A_0(m_Z^2) + 
\frac{(s_W^2 - c_W^2)^2}{c_W^2s_W^2}(m_Z^2 - 2A_0(m_Z^2)))
\\
& & \Sigma_{VS}^{H^{\pm}H^{\pm}}(q^2)  = -\frac{\alpha}{16{\pi}s_W^2}
(4s_W^2((m_{H^{\pm}}^2+ q^2)B_0(q^2, 0, m_{H^{\pm}}^2)-2q^2B_1(q^2, m_{H^{\pm}}^2, 0))
\nonumber\\ && \qquad\qquad\quad
+\frac{(c_W^2 - s_W^2)^2}{c_W^2}(A_0(m_Z^2) + (m_{H^{\pm}}^2 +q^2)B_0(q^2, m_{H^{\pm}}^2, m_Z^2) 
\nonumber\\ & & \qquad\qquad\quad - 2q^2B_1(q^2, m_{H^{\pm}}^2, m_Z^2)) +(A_0(m_A^2) + 
(m_W^2+ 4q^2)B_0(q^2, m_A^2, m_W^2)\nonumber\\ && 
\qquad\qquad \quad +4q^2B_1(q^2, m_W^2, m_A^2)) +{c_{\beta\alpha}^-}^2(A_0(m_h^2) + 
(m_W^2+ 4q^2)B_0(q^2, m_h^2, m_W^2) 
\nonumber\\ && \qquad\qquad\quad + 4q^2B_1(q^2, m_W^2, m_h^2)) +{s_{\beta\alpha}^-}^2(A_0(m_H^2) + 
(m_W^2+ 4q^2)B_0(q^2, m_H^2, m_W^2) \nonumber\\ && \qquad\qquad \quad
+ 4q^2B_1(q^2, m_W^2, m_H^2)))
\end{eqnarray}
%\end{document}